# Prototyping and Evaluation of Infrastructure-assisted Transition of Control for Cooperative Automated Vehicles

Baldomero Coll-Perales*, Joschua Schulte-Tigges[γ], Michele Rondinone[Ψ], Javier Gozalvez*, Michael Reke[γ], Dominik Matheis[Ψ], Thomas Walter[Ψ]

*Abstract*— **Automated driving is now possible in diverse road and traffic conditions. However, there are still situations that automated vehicles cannot handle safely and efficiently. In this case, a Transition of Control (ToC) is necessary so that the driver takes control of the driving. Executing a ToC requires the driver to get full situation awareness of the driving environment. If the driver fails to get back the control in a limited time, a Minimum Risk Maneuver (MRM) is executed to bring the vehicle into a safe state (e.g., decelerating to full stop). The execution of ToCs requires some time and can cause traffic disruption and safety risks that increase if several vehicles execute ToCs/MRMs at similar times and in the same area. This study proposes to use novel C-ITS traffic management measures where the infrastructure exploits V2X communications to assist Connected and Automated Vehicles (CAVs) in the execution of ToCs. The infrastructure can suggest a spatial distribution of ToCs, and inform vehicles of the locations where they could execute a safe stop in case of MRM. This paper reports the first field operational tests that validate the feasibility and quantify the benefits of the proposed infrastructure-assisted ToC and MRM management. The paper also presents the CAV and roadside infrastructure prototypes implemented and used in the trials. The conducted field trials demonstrate that infrastructure-assisted traffic management solutions can reduce safety risks and traffic disruptions.**

***Index Terms*— Automated driving; automated vehicles; connected automated vehicles; CAV; experimental evaluation; field tests; Minimum Risk Maneuver; MRM; prototype; transition of control; ToC; take over request; traffic management; V2X.**

## I. Introduction

Automated Driving (AD) capabilities are continuously increasing thanks to advances in perception, planning and control. Highly and fully automated driving has been piloted and showcased in different operational design domains and their market deployment is not far away [1]. However, different studies have shown that AD will not be always possible and [2] lists a non-exhaustive number of situations where AD is challenged. These situations might be caused by static or dynamic factors (e.g., roadworks) that alter usual road infrastructure layouts (e.g., road markings and signs), adverse weather conditions (e.g., snow and fog), or blocking elements (e.g., vehicles). These situations can require a Transition of Control (ToC) to manual driving if Automated Vehicles (AVs) reach their functional limits. In case of ToC, the driver needs some time to acquire full situational awareness and safely resume the driving tasks. Previous studies have shown that drivers taking over from high levels of automation experience a phase of reduced driving performance that can result in irregular and erratic behaviors [3]. This includes difficulties to keep the vehicle in the lane [4][5] and a tendency to brake less precisely or over-brake [6]. If the driver does not respond to a ToC request, an AV shall execute a so-called Minimum Risk Maneuver (MRM) to bring the vehicle into a safe state. An MRM might consist in decelerating to full stop or change lane to occupy a safe spot [7]. ToCs can generate safety risks if not executed properly. They can also imply risks for surrounding traffic participants and disrupt the traffic flow. This is especially the case in so-called Transition Areas where multiple AVs can execute ToCs simultaneously.

Fig. 1 illustrates this scenario in a road section where AD is not possible or not allowed (no AD zone in the following)[1]. In Fig. 1, two AVs approach the no AD zone. The AVs perform a ToC and request the driver to take over control (Take Over Request –TOR– in the following) at the same time ($t_1$ in Fig. 1) just before entering the Transition Area. Right after the ToC ($t_2$ in Fig. 1), the vehicles will be closer to the roadworks and a maneuver is necessary for the vehicle on the blocked lane to move to the free lane. This can imply a safety risk and traffic disruption, for example, if the traffic is dense or both vehicles are close to the roadworks and have little time and space to maneuver. The example illustrated in Fig. 1 has considered that both drivers resume driving manually after the TOR. If this is not the case, this would have led to the execution of an MRM with the associate potential risk of blocking the traffic in case any of the vehicles would stop on the driving lane. Fig. 1 helps illustrate the challenges and risks derived from the uncoordinated execution of multiple ToCs.

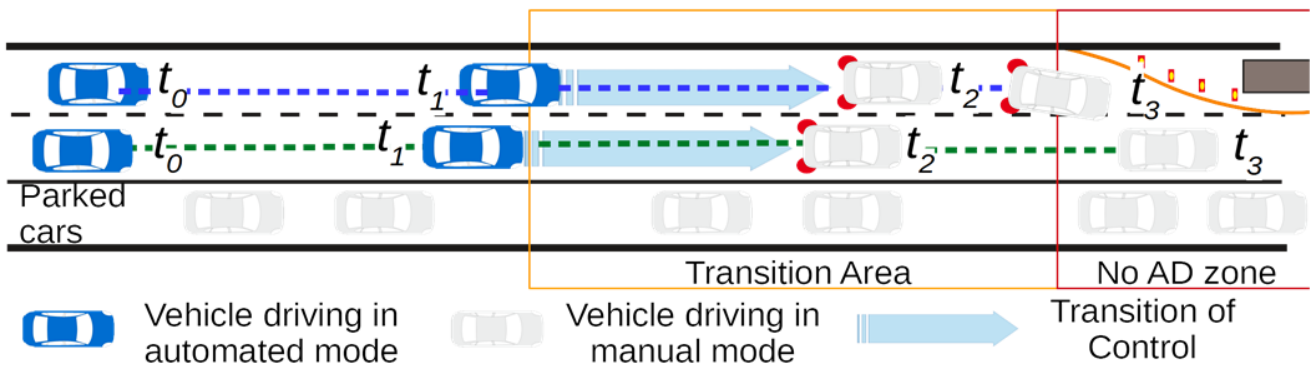

Fig. 1. Effects of multiple ToCs.

*(Corresponding author: Baldomero Coll-Perales.)*

This work was supported in part by the European Commission through the H2020 TransAID project (Grant Agreement no. 723390), the Generalitat Valenciana (BEST/2019/039, GV-2021), and the Spanish Ministry of Science, Innovation and Universities, AEI, and FEDER funds (TEC2017-88612-R, IJC2018-036862-I).

Baldomero Coll-Perales and Javier Gozalvez are with Universidad Miguel Hernandez, Elche, 03202, ES (e-mail: {bcoll, j.gozalvez}@umh.es).

Joschua Schulte-Tigges and Michael Reke are with FH Aachen, Aachen, 52066, DE (e-mail: joschua.schulte-tigges, reke@fh-aachen.de).

Michele Rondinone, Dominik Matheis and Thomas Walter are with Hyundai Motor Europe Technical Center, 65428, Rüsselsheim, DE (e-mail: {MRondinone, dmatheis, TWalter}@hyundai-europe.com).

[1] The use of roadworks to justify that there exists a no AD zone is only illustrative. A no AD zone can be due to multiple factors.

Traffic management centers and road infrastructure operators can play a key role in mitigating the negative effects of ToCs. To this aim, they can exploit the advent of C-ITS technologies and vehicle connectivity to manage and coordinate the execution of ToCs. The infrastructure can use different ITS and C-ITS technologies for an extended awareness and more complete knowledge of the driving environment. It can also use Infrastructure-to-Vehicle (I2V) and Vehicle-to-Infrastructure (V2I) communications to exchange information with vehicles and assist them in the management and execution of their ToCs. In this context, the EU-funded H2020 TransAID project proposes the development of infrastructure-assisted or I2V-assisted traffic management measures to manage the execution of ToCs, in particular under Transitions Areas. The project defines different, but complementary, classes of traffic management measures when the need for a ToC can be detected in advance [2]:

- *Prevent ToC*: the infrastructure provides supporting information (e.g., alternative driving paths, or speed, lane change advices) to Connected and Automated Vehicles (CAVs) that allow them to prevent a ToC and maintain their automated driving level.
- *Distribute ToC*: the infrastructure provides CAVs with suggestions to execute ToCs at different times or locations to distribute ToCs and reduce the potential negative effects derived from the simultaneous execution of multiple ToCs.
- *Manage ToC:* the infrastructure provides CAVs with suggestions to better manage the ToC, e.g., a safe spot to stop at in case of MRM.

Simulation studies have demonstrated the potential of these measures in improving traffic efficiency and safety. For example, [8]-[10] showed that an adequate distribution of ToCs can increase the throughput by 20% and the average network speed by 120% while reducing the safety risks. In particular, [8] and [10] showed that the distribution of ToCs reduces the probability of experiencing a time to collision lower than 3s by 88%. These simulation studies pave the way for the deployment of advanced measures to manage potential ToCs of CAVs. However, experimental trials are necessary to test the effectiveness of these measures under real world conditions and considering all implementation factors that are difficult to model in simulations. Prototype implementations and experimental proof of concept are necessary to have tangible and first-hand indications on the feasibility and quality of the implemented solutions, especially in terms of reliability, robustness and user acceptance in real-world conditions.

In this context, this paper presents, to the best of the authors' knowledge, the first prototype implementation and demonstration of infrastructure-assisted ToC management measures for CAVs. The implementation and demonstration are based on the TransAID's proposals to distribute and manage ToCs before a no AD zone. The infrastructure assists CAVs with indications on where and when to execute a ToC, and with information about the presence and location of safe spots where to park in case of MRM. The functionalities needed to experimentally validate and demonstrate these ToC management measures have been implemented on CAV and roadside infrastructure prototypes. These prototypes include the capacity to exchange information to manage and coordinate maneuvers using V2X communications. The implemented prototypes are used to demonstrate the feasibility and effectiveness of TransAID's *Distribute ToC* and *Manage ToC* traffic management measures, as well as their advantages compared to a baseline scheme where CAVs are only informed about the presence and location of a critical situation downstream. The experimental field trials show that the ToC management measures effectively prevent multiple CAVs from executing ToCs at close by locations. These measures also support CAVs much better in finding a safe spot for parking in case of MRM. The study complements the experimental field trials with additional numerical evaluations to analyze the ToC management measures under a larger range of settings and conditions.

## II. Related Work

Recent studies have shown that ToCs can negatively impact the traffic efficiency and safety [8]-[10]. This is a consequence of the driver's (lack of) ability to respond to the TOR issued by AVs [11][12]. For example, [12] analyzes the time it takes to a driver to successfully resume control from an AV. The study shows that when ToCs are more predictable and performed in regular intervals, drivers' lateral control of driving and steering correction are stable after a lag time of around 10s. However, when the transition from automated to manual driving happened in variable intervals, it can take drivers around 35s-40s to stabilize the lateral control of the vehicle. According to [3], the time needed by a driver to resume stable driving is also influenced by the level of driving automation and by the traffic density. Drivers taking over from high levels of automation experience a phase of reduced driving performance that results in irregular and erratic behaviors [3]. The study in [13] shows that the impact of ToCs on traffic operations will be amplified by the presence of mixed traffic where manual and AVs coexist because of their complex interactions. In addition, [13] demonstrates through simulations that mixed traffic can induce severe congestions around roadworks areas in the absence of an adequate traffic management.

The infrastructure can help manage traffic and prevent or mitigate some of the negative effects of ToCs. In fact, the European Road Transport Research Advisory Council (ERTRAC) highlights the importance and role of the infrastructure for the development of AD (especially for the higher automation levels) [14]. ERTRAC identifies in [14] various Infrastructure Support levels for Automated Driving (ISAD) levels based on the capability of the infrastructure to assist CAVs on certain road segments. In particular, ERTRAC defines five ISAD levels ranging from "E. Conventional infrastructure/No AV support" to "A. Cooperative driving". Under ISAD level A, the infrastructure should be able to assist CAVs in order to optimize the overall traffic flow. This study focuses on ISAD level A following ERTRAC's classification.

The report in [2] describes how an infrastructure (ISAD level A) that provides speed, gap, lane advices or alternative paths advices can prevent CAVs from executing ToCs in some traffic situations. This could be the case, for example, in a road segment where a lane is blocked/closed (e.g., by an obstacle, roadworks, etc.) and vehicles need to identify alternative routes to overpass it. Another scenario highlighted in [2] where the support from the infrastructure can help prevent ToCs is a highway merge segment. In this case, the infrastructure can provide speed and lane advices to create the necessary gaps on the right-most mainline to facilitate the merging of the CAVs coming from the on-ramp. The CONCORDA project also analyzed other examples and scenarios where the infrastructure can support CAVs in handling traffic situations that might lead to a ToC. This includes, for example, the use of the roadside infrastructure to support ToCs in a 'highway chauffeur' use-case [15]. The project proposes infrastructure-assisted solutions that rely on the V2X transmission of ETSI-based DENM messages [16]. The infrastructure uses these messages to warn CAVs about hazard situations occurring downstream and to notify CAVs that they should reduce the speed and/or request the driver to take over control if the AD system cannot handle the situation.

First studies have analyzed how infrastructure-assisted traffic management policies using I2V communications can help manage traffic and ToCs. This includes the study in [13] that shows how these policies can improve the traffic's efficiency in the proximity of roadworks even with high traffic intensity. The study in [8] shows how the infrastructure can help coordinate and distribute ToCs in order to improve the traffic efficiency and safety in areas where a large number of ToCs may be expected. These studies provide important insights into how infrastructure-assisted traffic management solutions can help improve automated driving. However, they are based on simulations only, and field trials are necessary to validate the potential of infrastructure-assisted traffic management solutions and evaluate their feasibility under real-world conditions. In this context, this study presents the first prototype and field trials that evaluate the effectiveness of infrastructure-assisted solutions to manage ToCs and MRMs in automated driving.

## III. Infrastructure-Assisted ToC/MRM Management

The scenario selected in this study for the implementation and evaluation of infrastructure-assisted ToC management is a road section that has a no AD zone at the end. Without loss of generality, this no AD zone is caused by the presence of roadworks that block a driving lane. The road infrastructure and the vehicles in the scenario are equipped with V2X technologies so that they can exchange the messages necessary to implement the infrastructure-assisted ToC management measures. The study implements and trials two infrastructure-assisted ToC management schemes. The first one is a baseline solution based on the proposal developed by the CONCORDA project and that relies on the exchange of DENM messages. The second one is a proposal co-developed by the authors under the TransAID project that relies on the exchange of Maneuver Coordination Messages (MCM) [17][18] for distributing ToCs in time and space and managing MRMs.

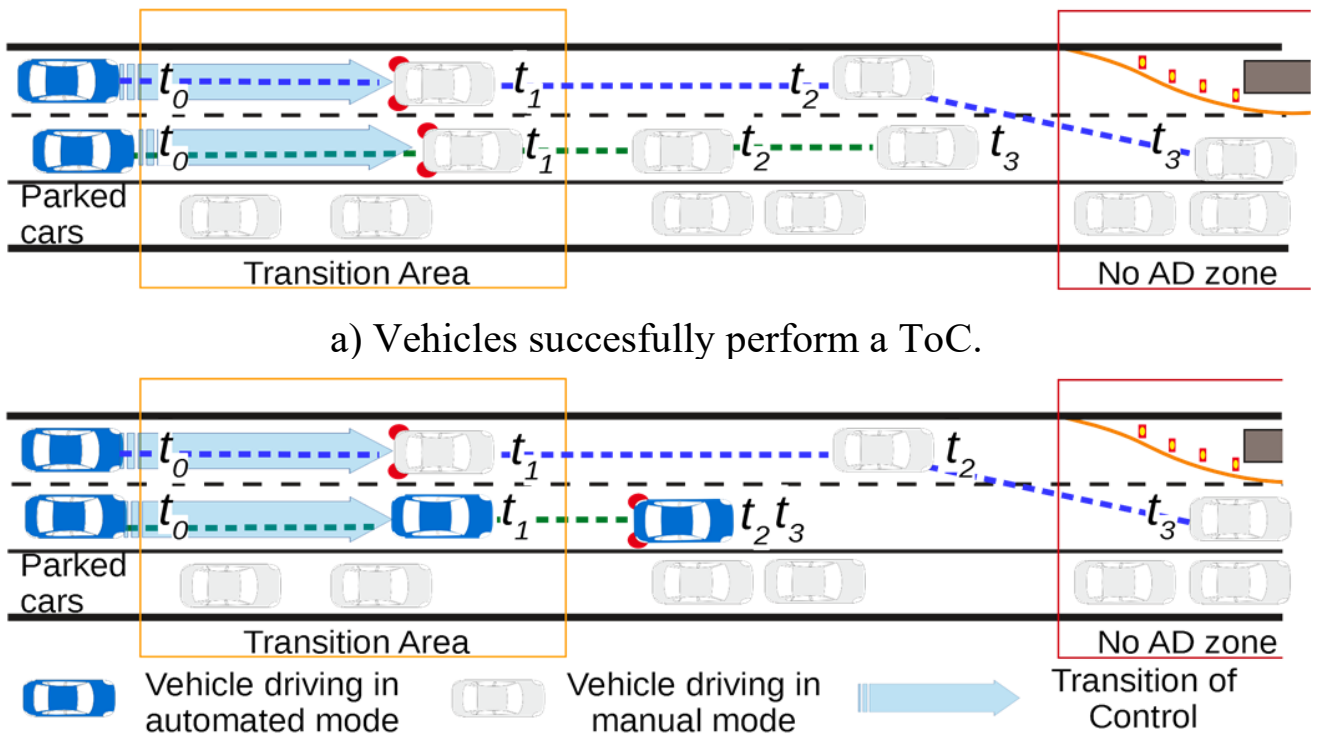

a) Vehicles succesfully perform a ToC.

b) Execution of an MRM.

Fig. 2. DENM-based infrastructure-assisted ToC management.

The baseline infrastructure-assisted ToC management scheme is based on the solution developed under the CONCORDA project. To the best of the authors' knowledge, this is the only infrastructure-assisted ToC management proposal available in the literature prior to the proposals developed under TransAID. The infrastructure notifies oncoming CAVs about the presence of the no AD zone downstream using DENM messages. DENMs are periodic broadcast messages used to inform recipient vehicles about the occurrence of a hazardous event on the road. A DENM contains information about the hazard type (roadworks in this scenario), its location as well as the relevance distance or distance from which oncoming vehicles shall consider the information relevant. If a CAV driving towards a no AD zone receives a DENM, the CAV issues a TOR when it is at a distance to the no AD zone equal to the relevance distance. Fig. 2 illustrates how this scheme shifts the Transition Area giving the drivers more time to coordinate before reaching the no AD zone. The relevance distance is a fix value that applies equally to all vehicles. In this case, the two vehicles approaching the no AD zone would issue the TORs and execute the ToCs at approximately the same location ($t_1$ in Fig. 2.a). However, the impact of concurrent ToCs is less negative compared to the scenario in Fig. 1, where there is no infrastructure support, since drivers have more time to coordinate their maneuvers before reaching the no AD zone. However, some risks can still occur if nearby vehicles execute ToCs at the same time, since drivers need some time to control adequately the vehicle after a period of inactivity. If the driver fails to respond in time to the TOR (CAV on the right lane at $t_1$ in Fig. 2.b), an MRM must be executed, and the CAV must stop. Fig. 2.b represents a scenario in which the CAV has no other option than stopping in the driving lane since there are parked cars. Stopping a CAV on the driving lane can block traffic and generate significant traffic risks.

The DENM-based ToC management scheme triggers the execution of ToCs at a given location for all CAVs and can turn out to be inefficient. More efficient traffic measures should allow the possibility to suggest actions individually to CAVs based on their specific context and with the objective to ensure a global positive impact on the traffic flow and safety. To this aim, this study implements a second and more advanced infrastructure-assisted ToC management scheme aimed at preventing the negative effects of ToCs on the traffic flow and safety. This second scheme relies on the MCMs messages under definition in ETSI for cooperative maneuvering, and that [19][20] extends to introduce the possibility for the infrastructure to support cooperative maneuvers. These extensions enable the road infrastructure to send individual advices to the CAVs on how to manage ToC and safe spots (among others) in order to increase the overall traffic safety and efficiency. Details about the implementation of the MCM messages are provided in Section IV.A.

In the MCM-based ToC management scheme, the infrastructure not only notifies of an incoming ToC but also suggests a spatial distribution of ToCs over a wider Transition Area. Fig. 3.a illustrates the operation of this second scheme. In the example, the road infrastructure suggests the two close-by driving CAVs to trigger the ToC at two different locations, e.g., at their locations at $t_0$ and $t_1$. This results in that when the driver of the vehicle on the left lane takes over ($t_2$ in Fig. 3.a), the driver of the vehicle on the right lane has already recovered her/his driving skills, is more attentive and can better handle possible surrounding safety risks. At $t_3$, the two drivers can effectively coordinate their actions and pass through the roadworks without disrupting the traffic flow. The MCM-based ToC management scheme also implements a procedure to handle efficiently MRMs. This procedure constantly suggests CAVs (based on their location) road sections with safe spots where the vehicle could stop safely and automatically if drivers fail to take over. Fig. 3.b illustrates this scenario and considers the case in which the driver of the CAV on the right lane is not responding to the TOR ($t_1$ in Fig. 3.b). In this case, the MCM-based ToC management scheme will inform the vehicle of a safe spot location. This prevents the vehicle from stopping on the driving lane. With this information, the CAV will implement an MRM guiding to a free section of the parking lane. This prevents risks and blockage of the driving lanes.

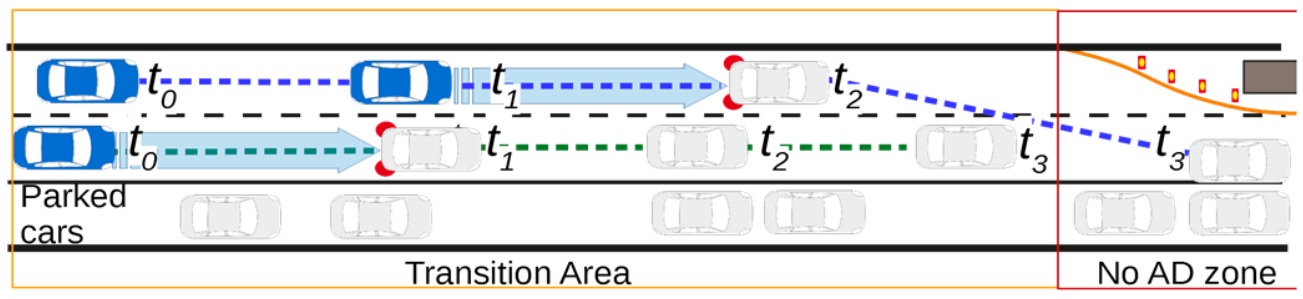

a) Distribution of ToC.

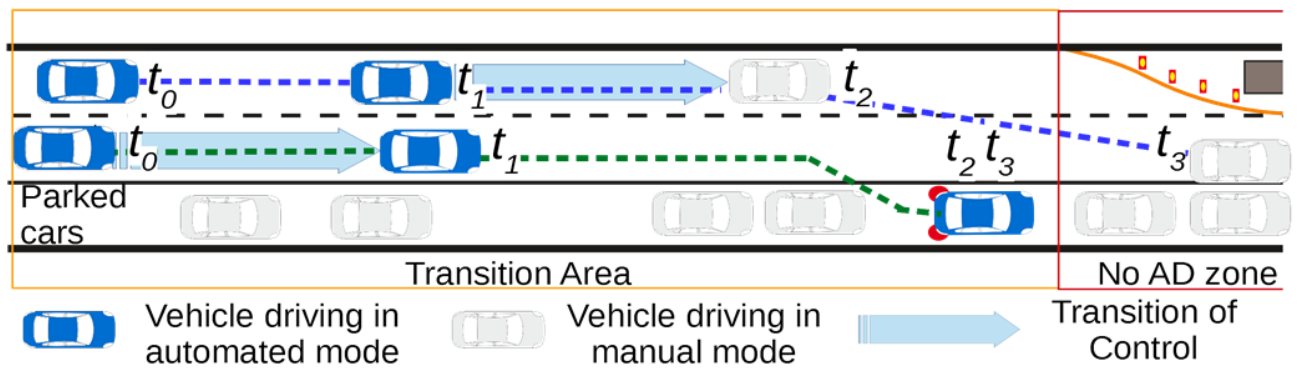

b) MRM execution guiding the CAV to a safe spot.

Fig. 3. MCM-based infrastructure-assisted ToC management.

## IV. Prototype Platform

The prototype platform implemented in this study includes a Road Side Unit (RSU) and a CAV (Fig. 4). The logical architecture of the platform is depicted in Fig. 5. The RSU and CAV are equipped with a V2X module that enables V2I communications. The RSU uses the information received in V2X messages and the contextual information collected from the Road Side Sensors module as input to the Traffic Monitoring module. The Traffic Management module uses this information to implement infrastructure-assisted ToC management measures. The RSU transmits these ToC management measures to the CAV via the V2X module. The CAV complements the information it collects from the In-vehicle Sensors module with the V2X messages received from the RSU. The Autonomous Driving Software (AD SW) module uses as inputs the information received from the V2X and In-vehicle Sensors modules to plan and execute the CAV's autonomous maneuvers. In addition, the AD SW module provides information to the V2X module to create the V2X messages to be transmitted. The developed platform also includes a Human-Machine Interface (HMI) that acts as an interface between the AD SW module and the driver.

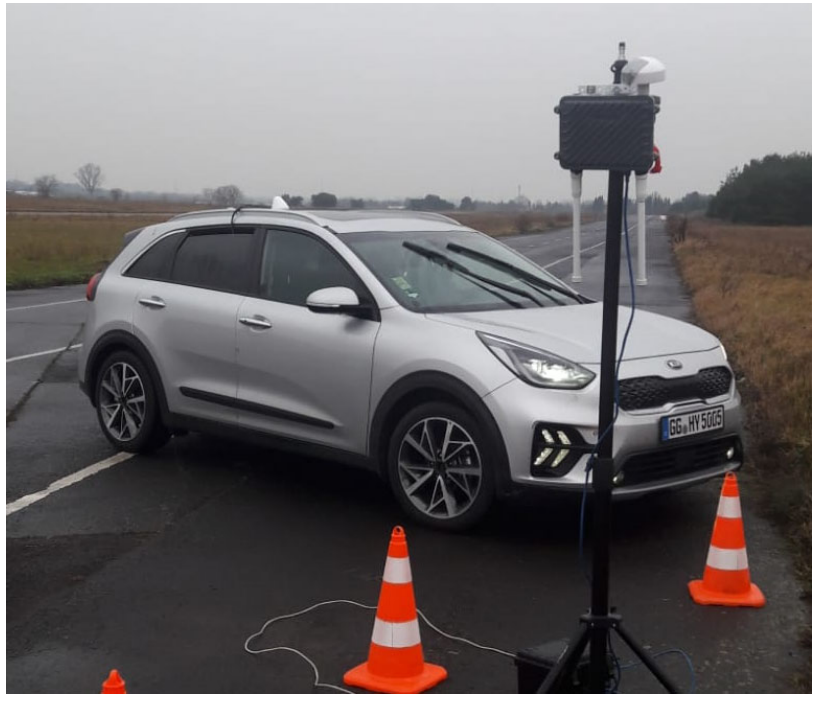

Fig. 4. RSU and CAV prototypes.

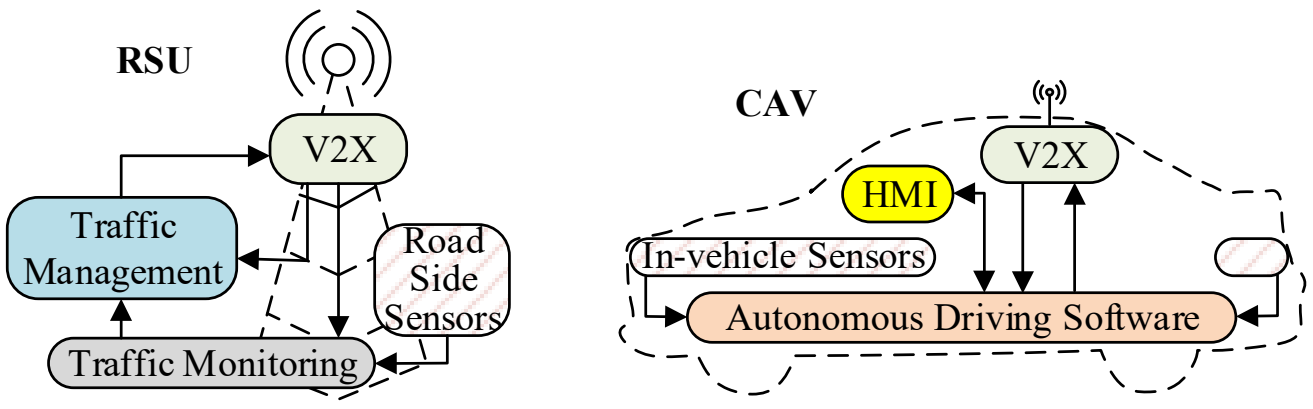

Fig. 5. Logical architecture of the cooperative V2X maneuvering platform.

### A. *V2X messages extension*

The implementation of advanced infrastructure-assisted ToC management measures has required extending some currently available standard V2X messages. In particular, extensions are made to the ETSI ITS CAM [21] and MCM messages [20]. The implemented extensions ensure backwards compatibility and interoperability with the original V2X messages. In addition, the extensions complement the original V2X messages while maintaining the logic and coherence of their contents.

ETSI ITS CAM messages provide information about position, dynamics and basic attributes of the transmitting station [21]. The structure of CAM messages is made of different containers. For example, a CAM transmitted by a vehicle shall comprise one basic container, one high frequency container and may also include a low frequency container and one or more special containers. The basic container includes basic information related to the transmitting station, e.g., position and type of station. The high frequency container includes information that shall be transmitted with high frequency like vehicle's speed and acceleration. The low frequency container includes static or slow-changing vehicle data like the status of the exterior lights or the vehicle role (i.e., public transport, emergency vehicle, etc.). The special vehicle container includes information specific to the vehicle's role. It should be noted that the infrastructure can track the movement of CAVs using the information in CAMs messages. However, this information cannot be used by the infrastructure to assess the actual traffic demands and compositions (i.e., mix of conventional, semi-automated and AVs) since CAMs do not include currently information about the automated level. This is important to derive traffic management decisions and individual suggestions for CAVs. To fill these gaps the CAM originated by a vehicle has been extended to include an *AutomatedVehicle* container that CAVs use to notify their current SAE automation level [22]. Similar extensions are proposed in EU funded projects like MAVEN [23].

The ETSI Technical Committee on ITS is currently defining the Maneuver Coordination Service to implement cooperative maneuvering or cooperative driving. This includes the definition of MCM messages. Current MCM message proposals allow vehicles to exchange information about their planned and desired trajectories. With this information and the right of way driving rules, vehicles can safely coordinate their maneuvers. The first version of MCM messages focuses on maneuver coordination between vehicles using V2V communications (ETSI has recently started to investigate the role of the infrastructure in the Maneuver Coordination Service [19]). Extensions to the MCM have been proposed by the authors in [19] and [20] to include the possibility for the infrastructure to support the maneuver coordination with advices and suggestions. These extensions have been implemented in the prototype testbed and allows the RSU to send individual advices to the CAVs on how to handle a ToC, create gaps, change lane, set a target speed and address a safe spot. All these advices are aimed at supporting CAVs and increase the overall traffic safety and efficiency. Fig. 6 shows the format of the extended MCM. It includes common headers such as the *ItsPduHeader*, *GenerationDeltaTime* and *Basic Container* that are used to indicate the ID of the message and station, the generation time, and the reference position and type of station generating the message, respectively. The originating station can be either a vehicle or an RSU. The *Maneuver Container* includes a *VehicleManeuverContainer* if the MCM is transmitted by a vehicle or a *RSUSuggested ManeuverContainer* if it is transmitted by the RSU. The *RSU SuggestedManeuverContainer* includes the above-mentioned advices while the *VehicleManeuverContainer* includes information about the trajectories (planned and desired) to coordinate the maneuvers. The *VehicleManeuverContainer* could also include a response list (*AdviceResponseList*) to the advices that the CAV has received from the RSU. This response list is used to acknowledge whether the CAV is following the advice suggested by the RSU or not.

<table>
<tr><td rowspan="5">MCM</td><td colspan="4">ItsPduHeader</td></tr>
<tr><td rowspan="4">ManeuverCoordination</td><td colspan="3">GenerationDeltaTime</td></tr>
<tr><td rowspan="3">MCMParameters</td><td colspan="2">BasicContainer (ReferencePosition + StationType)</td></tr>
<tr><td rowspan="2">ManeuverContainer = CHOICE</td><td>VehicleManeuverContainer (Dynamics + plannedTrajectory + desiredTrajectory + AdviceResponseList)</td></tr>
<tr><td>RsuSuggestedManeuverContainer (list target vehicle-specific advices: transition of control, gap, lane change, speed, safe spot)</td></tr>
</table>

Fig. 6. MCM format [20].

For the purpose of the tests conducted in this work, the following containers are of special interest:

1) *RSUSuggestedManeuverContainer:*

*TransitionOfControl*: this is an advice transmitted from the RSU to the CAV about how to handle the ToC. The advice is identified by a unique ID (*AdviceID*) that the CAV can use to provide feedbacks to the RSU. The RSU can also indicate to the CAV what automation level it should adopt after the ToC. Finally, the RSU advices the CAV where/when it should perform the ToC. This is indicated through either the *Transition AdviceDistanceRange* or *TransitionAdviceTimeWindow*. These parameters include a range of locations or time window, respectively, where the ToC should be performed.

*SafeSpot*: this advice is used by the RSU to inform the CAV about a location where a safe spot is available to do a safe stop. The advice is identified by a unique ID (*AdviceID*). The advice also includes the *SafeSpotAdviceRange* parameter that includes the range of locations where the CAV can do the safe stop.

2) *VehicleManeuverContainer:*

*AdviceResponseList*: this is a list that can include feedbacks for any of the advices transmitted by the RSU to a CAV. The reported feedbacks also indicate at what level the CAV is following or not the received advice through a *Compliance Status* parameter. This parameter might indicate, for example, that the advice was received, and that the CAV will try its execution if the driving conditions allow it. Based on the feedback received, the RSU might decide to retransmit new advices for a specific CAV.

### B. RSU implementation

#### 1) V2X module

The V2X module is implemented using a Cohda Wireless's MK5 RSU (see Fig. 4). The MK5 RSU is compliant with the latest European ETSI C-ITS standards for V2X communications at the different layers of the OSI stack. The integrated RoadLINK chipset implements the ETSI ITS G5 Access layer that profiles at European level the IEEE 802.11p standard [24]. A specific Application layer has been developed in this study and added to the RSU V2X module. This Application layer implements the two infrastructure-assisted traffic management schemes and manages the transmission and reception of all V2X messages.

Fig. 7 shows the processing of V2X messages at the Application layer of the RSU's V2X module. On the reception path, V2X messages (e.g., MCM and CAM) arrive at the Application layer through a callback function (see 'Msg Callback' in Fig. 7). This callback function is invoked whenever any V2X message is received at the lower layers. V2X messages are identified at the Application layer by their unique port ID included in the Basic Transport Protocol (BTP) header. For example, standard V2X messages such as the CAM and DENM use the standard ETSI BTP port 2002 and 2001, respectively. The MCM's BTP port ID is not yet defined, and in the current implementation we use the port ID 2010. The V2X messages are identified after the 'BTP Header sorting' module. However, their content is not yet accessible because messages are encoded with Unsigned Packet Encoded Rules (UPER). The UPER decoding of the received V2X messages is performed utilizing as a base their Abstract Syntax Notation 1 (ASN.1) representation. For the case of the extended CAM and MCM messages, their ASN.1 representation is defined in [20]. When we decode a CAM, we forward the ID of the CAV that generated the message, its SAE level and its location information to the 'Traffic Management' and 'Traffic Monitoring' modules. To this aim, specific UDP interfaces are defined between the different modules. These interfaces are made of data structures which are populated using the information included in the V2X messages. For the decoded MCM messages, the data structures that are transmitted over the MCM UDP interface (see *IF1_MCM* in Fig. 7) are filled with the information included in the *VehicleManeuverContainer*. The *IF1_MCM* connects to the 'Traffic Management' module. The V2X module and 'Traffic Management' module are physically connected over Ethernet.

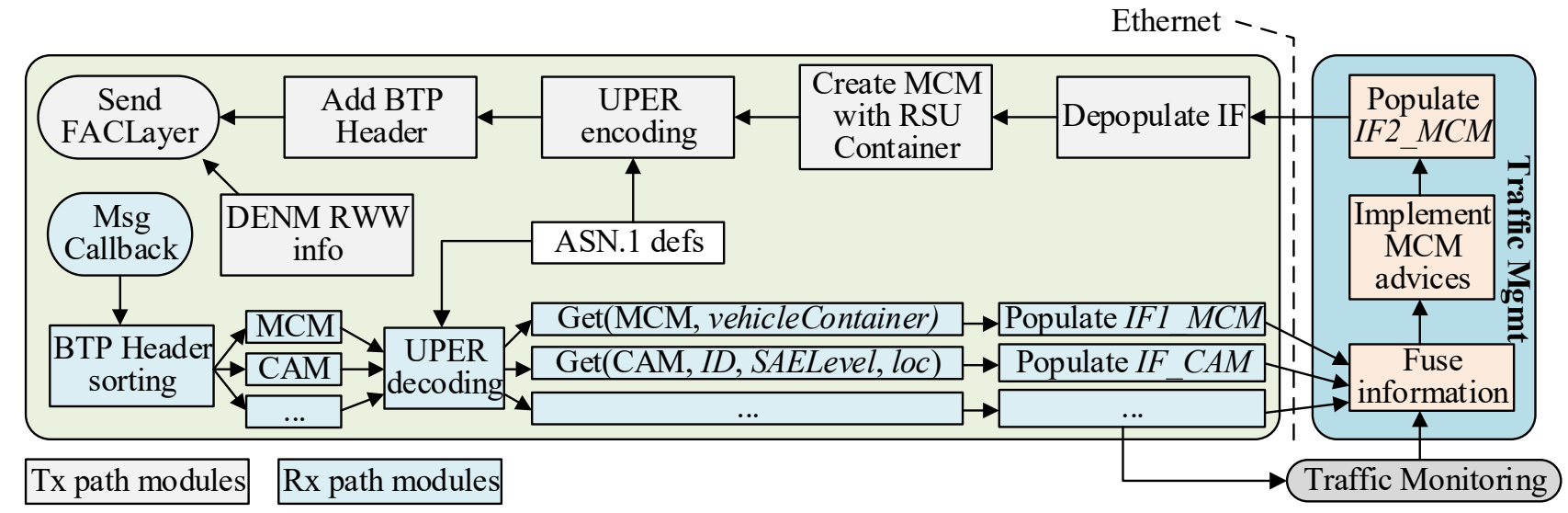

Fig. 7. V2X Application layer at the RSU.

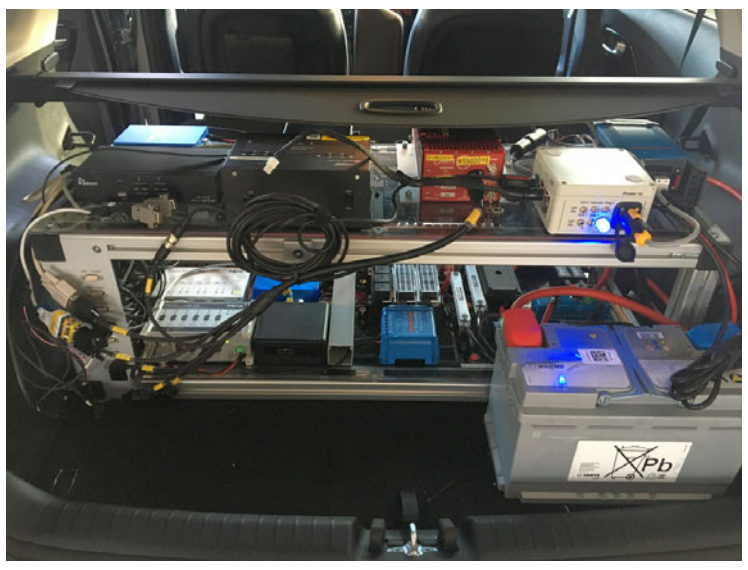

a) Automation system in the trunk

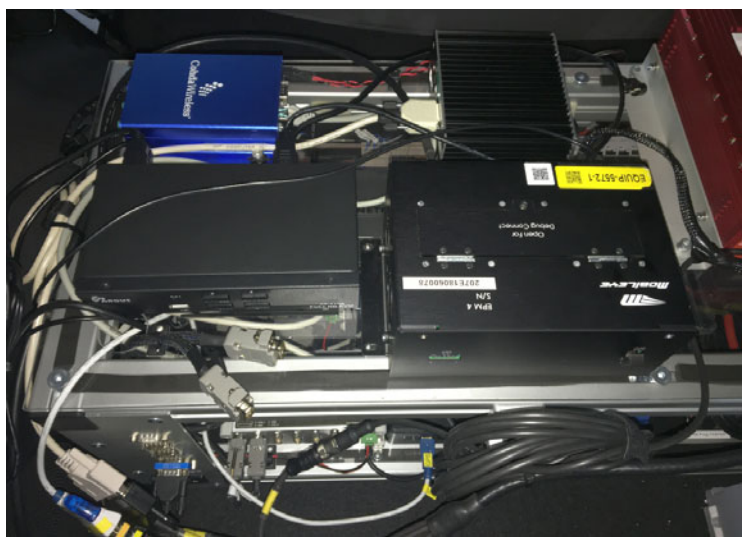

b) OBU and Mobileye PC

Fig. 8. Test vehicle with automation system in the trunk.

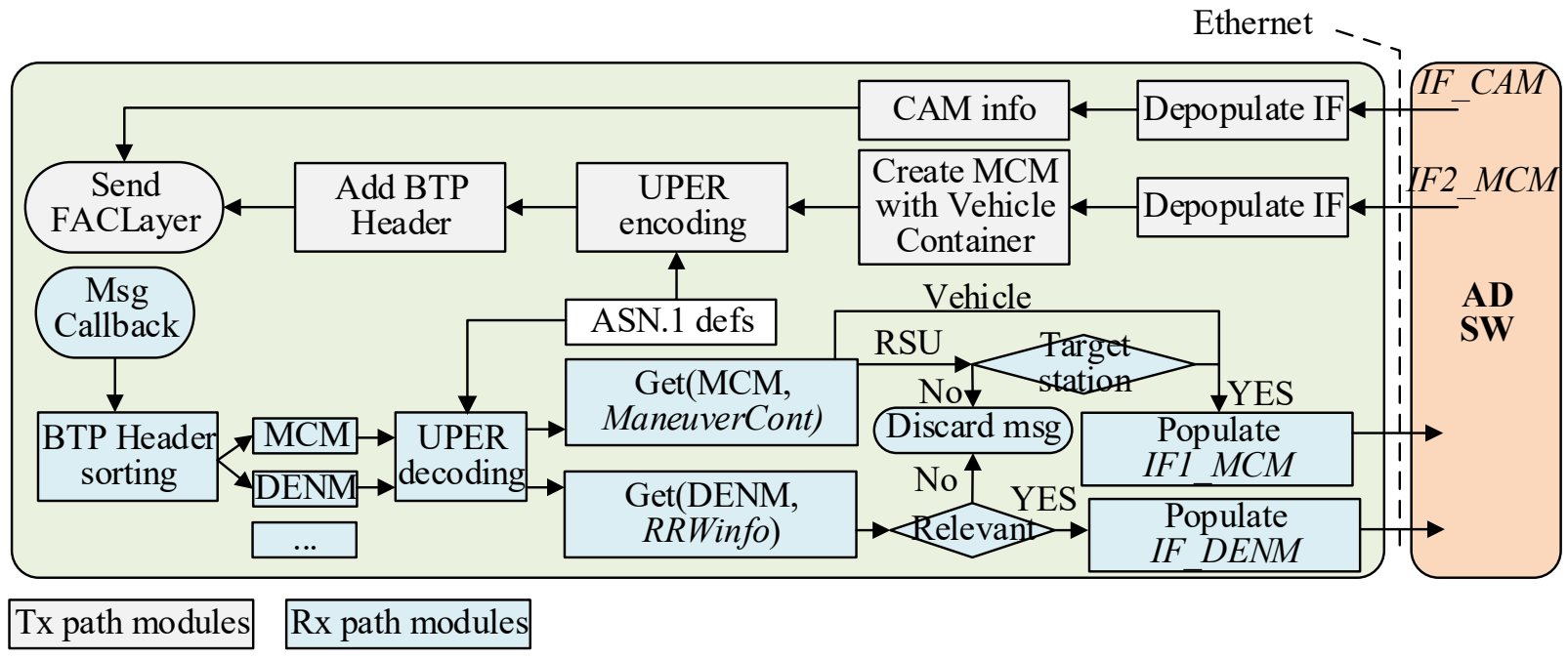

Fig. 9. V2X Application layer at the CAV.

On the transmission path, the V2X module's Application first extracts the data structures coming from the 'Traffic Management' module through the *IF2_MCM* (see 'Depopulate IF' module in Fig. 7). This information is used to create an MCM *RSUSuggestedManeuverContainer* ('Create MCM with RSU Container' module in Fig. 7). We also utilize the MCM ASN.1 representation to UPER encode the created MCM message ('UPER encoding' in Fig. 7). Finally, we add a BTP header to the message before sending it to the Facility Layer.

Fig. 7 also includes the 'DENM RWW info' module. This module is utilized in the implemented Application layer to configure the DENM messages that are periodically transmitted by the RSU to inform about the presence of the roadworks (RWW, RoadWorks Warning). In particular, the 'DENM RWW info' indicates the location of the roadworks, lane(s) affected, the relevance distance, etc. This information is sent to the Facility layer to generate the DENM message.

*2) Traffic Management and Monitoring*

Fig. 7 shows that the 'Traffic Management' module takes as inputs the *IF_CAM* and *IF1_MCM* with CAM- and MCM-related information, respectively, and additional information obtained from the 'Traffic Monitoring' module. This information is utilized to generate the MCM advices ('Implement MCM advices' in Fig. 7) that the RSU transmits using the V2X module. The information reported by the 'Traffic Monitoring' module can be utilized, for example, to identify the location of the safe spots. For this RSU prototype implementation, this information is considered available at this module even though the RSU is not equipped with the necessary sensors to detect this. We implement an UDP interface from the 'Traffic Management' module to the V2X module (*IF2_MCM* in Fig. 7) to send the MCM advices to the V2X module. The MCM *VehicleManeuverContainer* received at the 'Traffic Management' module can also include an *AdviceResponseList* (Fig. 6) that CAVs utilize to acknowledge the previously received advices from the RSU. If this is the case, this list is taken into account at the 'Implement MCM advices' module to remove the already acknowledged advices.

## *C. CAV implementation*

A KIA Niro (Fig. 4) was chosen as test vehicle. The vehicle has been equipped with a complete automation system in the trunk (Fig. 8.a). The On Board Unit (OBU) for V2X communications and the Mobileye camera system are integrated parts of this automation system (Fig. 8.b). The implemented CAV uses the Polysync DriveKit as the interface between the developed AD SW and the vehicle. Through the Polysync interface, it is possible to control the vehicle's acceleration, braking and steering via Controller Area Network (CAN) messages. The Polysync DriveKit also allows a safety driver to take back the control of the vehicle as soon as it presses a pedal or turns the steering wheel. The autonomous operations of the CAV are also subject to the information received through the V2X module and HMI. For the purpose of this study, the automated functionalities of the CAV prototype are not requested to cope with planning and control in reaction to surrounding objects' detection and tracking. Automated vehicle behavior in terms of ToC and MRM management was isolated from possible implications deriving from object detection.

*1) V2X module*

The V2X module at the CAV is implemented using a Cohda Wireless's MK5 OBU (Fig. 8.b). The main developments in the implemented CAV's V2X module have also focused on a specific Application layer. This Application layer manages the transmission and reception of all V2X messages that support the infrastructure-assisted ToC management measures.

Fig. 9 illustrates the implementation of the V2X Application layer at the CAV. On the reception path, the V2X module' Application processes the received V2X messages (e.g., MCMs or DENMs). For MCMs, the *ManeuverContainer* is accessed to identify whether the message was originated by an RSU or another CAV. If it was originated by the RSU, the *RSU SuggestedManeuverContainer* is analyzed to identify whether it includes advices addressed to the receiving CAV. If this is not the case, the MCM message is discarded. If there are advices addressed to the receiving vehicle, or if the MCM was originated by another CAV, the relevant information is transmitted through the *IF1_MCM* UDP interface that connects with the 'AD SW' module. When DENMs are received, the implemented Application accesses the RWW information and checks whether it is relevant to the CAV. If this is the case, the *IF_DENM* UDP interface is used to forward this information to the 'AD SW' module.

On the transmission path, the V2X Application layer receives information used to generate CAM and MCM messages. This information is transmitted by the AD SW using the *IF_CAM* and *IF2_MCM* interfaces. For example, the *IF_CAM* is used to transmit information obtained from the vehicle's CAN bus like speed, acceleration, heading and steering angle, as well as the currently operated SAE automation level. This information is used at the Facility layer to create the CAM containers and the *AutomatedVehicle* container (Section IV.A). The information necessary to generate MCMs (including its *VehicleManeuver Container*) is received through the *IF2_MCM* interface. *IF2_MCM* can also include feedbacks about the advices received from the RSU.

*2) In-vehicle sensors*

Testing the infrastructure-assisted ToC management schemes requires that the CAV prototype executes automated lateral and longitudinal control of the vehicle. To this aim, the CAV prototype mostly relies on the Mobileye EPM4 front-camera system as environmental sensing source. The Mobileye EPM4 is capable of processing the images captured of the road and transmits the processed data over the CAN bus. For example, the processing performed at the Mobileye EPM4 camera allows identifying the lane-marking of the road. For the detected lane-markings, the Mobileye EPM4 camera transmits information like distance to the lane-marking, curvature of the lane-marking and relative angle to the lane-marking. This information is made available in the CAN bus and is utilized by the AD SW to implement the vehicle lateral control.

*3) Automated driving software*

The AD SW installed in the CAV prototype is the ROS2-based platform for self-driving cars developed by FH Aachen in cooperation with HMETC [25]. The use of ROS (Robot-Operating-System ) for self-driving car platforms became very popular as it facilitates flexible modular designs that can be adapted to various types of vehicles. In addition, the second version of ROS (ROS2) supports real-time threads in their applications and real-time communication based on data distribution service. The platform presented in [25] has been extended and adapted to the necessities of the infrastructure-assisted ToC management testing.

Fig. 10 illustrates the main ROS2 nodes utilized for the execution of the automated maneuvers in the infrastructure-assisted ToC management testing. Fig. 10 also illustrates the messages exchanged between the ROS2 nodes and the topics they are subscribed to. First, let's analyze how the ROS2 nodes interact to achieve the longitudinal control. For longitudinal control, the vehicle must be able to adapt to a given speed at a given moment. In addition, it is also useful to control the acceleration and deceleration for an optimal driving behavior. This longitudinal control strategy allows a much smoother driving because the acceleration can be limited to prevent abrupt braking, for example, without losing the ability to perform an emergency brake. The control input values are desired speed, desired acceleration, current speed and current acceleration. Control outputs are percentage values of pedal position for both brake and throttle. For realizing such implementation of the longitudinal control, the 'ROS2CAR Vehicle Interface Node' receives the vehicle's current speed over the CAN bus and calculates the current acceleration. Both values are published on the /vehicle_state topic as vehicle_state_msg. A desired speed and desired acceleration can be set via ackermann_msg published over the /ackermann topic. The Vehicle Control Node Longitudinal subscribes to the /vehicle_state and /ackermann topics. Throttle and brake commands are periodically calculated by a proportional–integral–derivative controller and published on the topics /brake and /throttle as throttle_command_msg and brake_command_msg. 'ROS2CAR Vehicle Interface Node' subscribes to both (/brake and /throttle) and transmits the pedal positions to the Polysync DriveKit via the CAN bus.

For the lateral control of the vehicle, the main goal is to keep the vehicle in between the road's lane markings. To do so, the 'Camera Node' receives lane information from the Mobileye camera over the CAN bus. This information is published on the /camera topic periodically every 30ms. The 'Vehicle Control Node Lateral' subscribes to the /camera topic and calculates a lateral error to the center of the lane with the gathered information of both lane markings (left and right). With this lateral error, the steering wheel angle is controlled by a filtered proportional–integral–derivative controller. As a safety feature, the maximal allowed steering wheel angle is adapted to the vehicle speed (the higher the speed, the smaller the maximal steering wheel angle). The steering wheel angle is published as steering_command_msg over the /steering topic. The 'ROS2CAR Vehicle Interface Node' subscribes to the /steering topic and transmits it via the CAN bus to the Polysync DriveKit.

Besides longitudinal and lateral control, the AD SW must be also capable to execute maneuvers in reaction to received V2X information. The distributed ROS architecture helps with the implementation of this task. The 'V2X Node' passes the received information (for example, a ToC request) to the 'Mission & Maneuver Planning' node. This node then schedules a set of ToC related actions that depends on the information received and whether the driver reacts or not to the ToC request. The AD SW issues the TOR to the driver via the 'HMI Node' at the time indicated by the received information from the 'V2X Node'. If the driver does not react within a given time threshold, an MRM is executed. The AD SW is requested to coordinate different maneuvers for the execution of the MRM. This includes: speed adaptation to an objective MRM speed, lane change to the emergency lane, and stop in a safe spot. To coordinate these maneuvers, the 'Mission & Maneuver Planning' node can send out maneuver requests to the controlling nodes reusing the above mentioned longitudinal and lateral controllers.

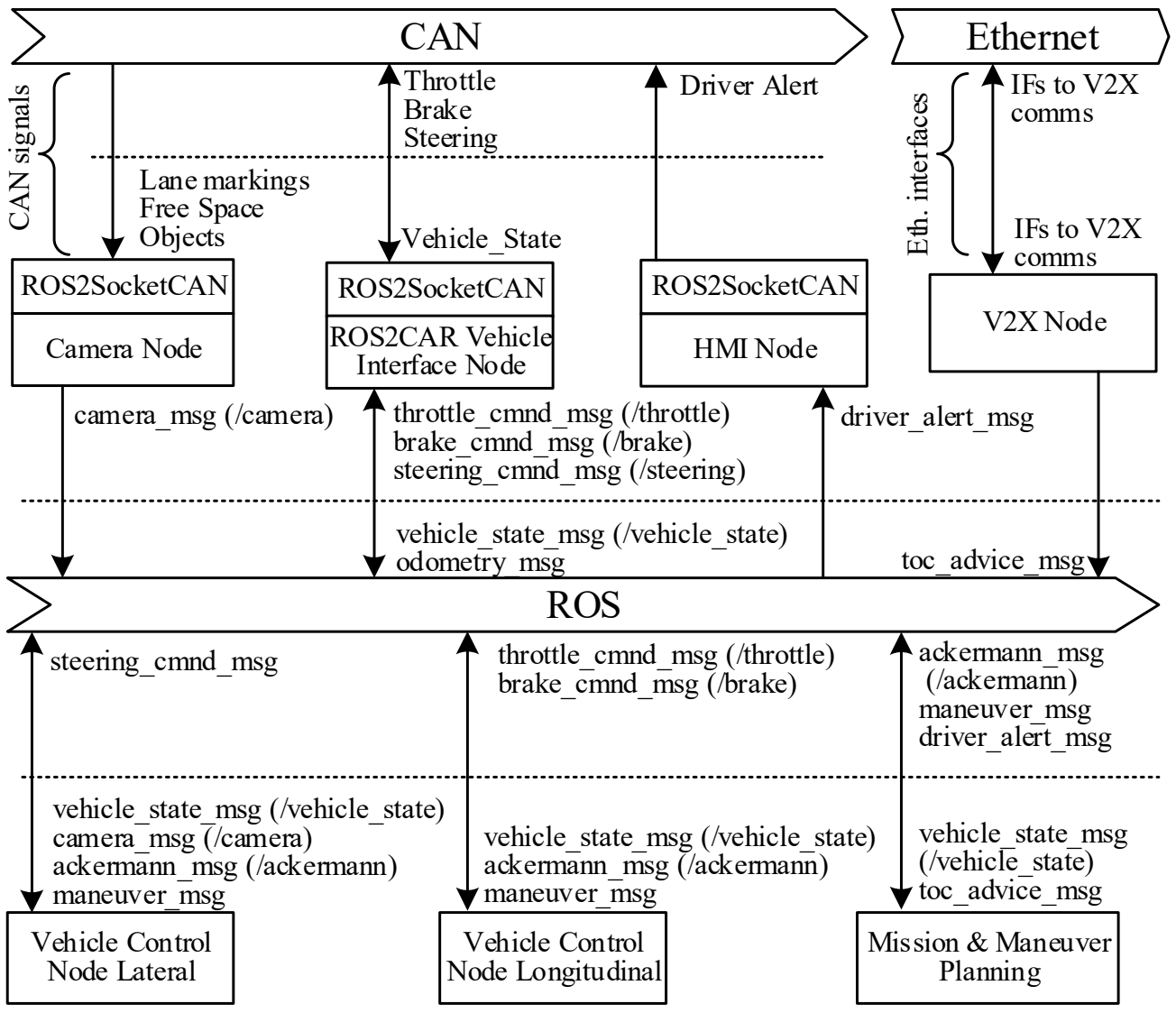

Fig. 10. ROS2 architecture of the AD SW functionalities in the CAV.

### 4) HMI

The CAV's prototype implementation includes a simple HMI that is used to inform the driver about the current and upcoming events. The AD SW runs an 'HMI Node' (Fig. 10) that handles the communication between the AD SW stack and the HMI hardware using a dedicated CAN message (driver_alert_msg). A ruggedized display was attached to the dashboard, which enables the test driver to quickly check the current system status. The display runs a small application. The application does not fulfill all rules and design guidelines of a series product but already addresses the need to avoid overloading the driver with information. The driver_alert_msg CAN message sent by the 'HMI Node' is received and processed internally by the application in the display to visualize the system status (e.g., TOR to the driver, MRM in execution) using various text messages, a countdown timer, and a progress bar. On top of the visualization, a system integrated beeper can be enabled increasing its beep frequency as the TOR's timeout approaches its expiration.

## V. Implementation of the Infrastructure-Assisted ToC/MRM Management

Field trials have been conducted at the proving ground of the Griesheim airport (Fig. 11). During the tests, the CAV uses the airport's main runway that has two lanes. The runway has a total length of approximately 1km, and it has been (virtually) divided into a 700-meter zone where the CAV can drive autonomously and a 300-meter zone where AD is not allowed (no AD zone in Fig. 11). The RSU is located at the start of the no AD zone. Fig. 11 shows the initial location of the CAV when the tests start. The CAV drives autonomously from this location and it reaches a target speed of 60 Km/h when it is 700m away to the no AD zone. The RSU informs the CAV that it should perform a ToC before reaching the no AD zone via DENM or MCM messages in the first and second infrastructure-assisted ToC management schemes, respectively. Safe spots to safely stop the CAV in case of MRM are available on the emergency lane set up next to the driving runway. These safe spots reflect, for example, free spaces between parked cars as indicated in Fig. 1-Fig. 3. For the sake of safety during the tests, safe spots are not obtained by parking real cars. Instead, the emergency lane is virtually divided into 25m-length sections that are randomly chosen as free or occupied in each test run. This random scenario configuration is made available to the CAV and RSU. A safe spot is made of 3 consecutive free sections that allow the CAV to safely perform a lane change from the driving runway to the emergency lane and stop in case of MRM. For each test run, at least a safe spot is available in the scenario. The scenario illustrated in Fig. 11 shows an example with one safe spot available at [75m, 150m]. It should be noted that the free section at [350, 375] would not be considered a safe spot to perform the lane change and stop since it is not long enough to safely do the MRM maneuver.

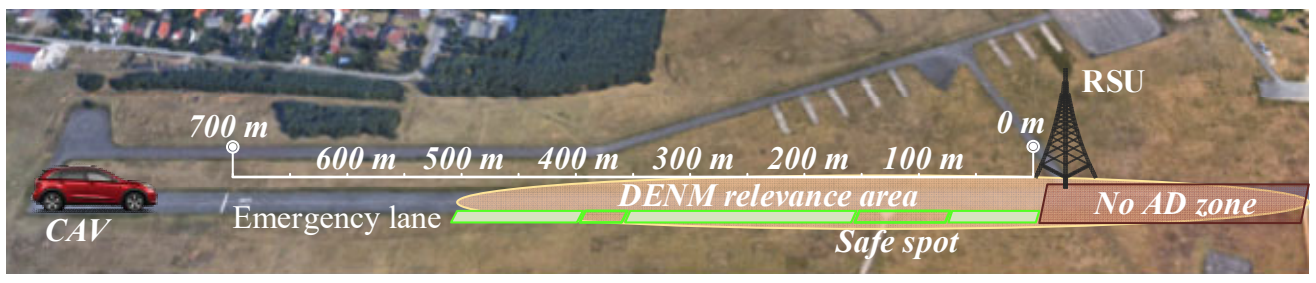

Fig. 11. Aerial view of the Griesheim airport facilities in Griesheim (DE).

We consider that from the moment a ToC is requested, the driver has a lead time $t_{TOR}$ of 10s to take over control before an MRM is executed[2]. This is independent of the infrastructure-assisted ToC management scheme under evaluation. During the TOR's lead time, the CAV continues driving at 60 Km/h. This study considers that the driver does not intervene in time to a TOR and the CAV always executes an MRM. This is to investigate the impact on the traffic safety and efficiency of the execution of MRM when it is triggered by a DENM-based or MCM-based ToC management schemes. Another common configuration for the DENM-based and MCM-based ToC management schemes during the MRM is that the parking maneuver is performed at $Speed_{MRM}$ that is set to 20Km/h.[3] The CAV must then slow-down from the driving speed (i.e., 60 Km/h) to $Speed_{MRM}$ before executing the parking maneuver.

### *A. DENM-based ToC management*

The RSU is configured to periodically transmit DENM messages at 1Hz to inform the approaching vehicles of the presence of the no AD zone. Even though the RSU's V2X range covers the whole testing area, DENM messages are only processed by the CAV when it reaches a DENM's relevance distance of 500m to the no AD zone[4] (Fig. 11). Fig. 12 shows the ToC processing at the AD SW for the DENM-based ToC management scheme. A TOR is triggered when the CAV enters the DENM's relevance area. During the TOR, the CAV continues driving at its current speed. If the TOR's lead time $t_{TOR}$ is consumed and the driver has not taken back control, the CAV slows down to $Speed_{MRM}$ as part of the MRM. With the information provided by the DENM, the CAV is not aware of the locations of the safe spots available to park. This study has considered three different variants of the DENM-based ToC management scheme that are designed to allow the CAV to search for a safe spot for a configured distance $d_{MRM}$. While searching, the CAV detects free or occupied sections on the emergency lane with its local sensors. It should be noted that the CAV searches for a safe spot without knowing whether one will be found or not. The three different variants of the DENM-based ToC management scheme implemented on the CAV are:

- DENM_$d_{MRM}$=0. When the CAV reaches the $Speed_{MRM}$, the AD SW is programmed to stop driving. In this case, the CAV only performs the parking maneuver (i.e., change to the emergency lane and stop) if a safe spot is available at the point the CAV reaches the $Speed_{MRM}$. Otherwise, the CAV stops on the driving lane.
- DENM_$d_{MRM}$=50. In this case, the CAV can search for a safe spot for 50m when it reaches $Speed_{MRM}$. When an MRM is executed, the CAV slows down to $Speed_{MRM}$ and drives at most 50 meters at $Speed_{MRM}$. If the CAV finds a safe spot while driving at $Speed_{MRM}$, it executes the parking maneuver. Otherwise, the CAV stops on the driving lane.
- DENM_$d_{MRM}$=unlimited. The CAV can search for a safe spot when it reaches the $Speed_{MRM}$ in all the relevance area. If it finds one, the CAV executes the parking maneuver. If the CAV does not find a safe spot, it stops on the driving lane just before the no AD zone.

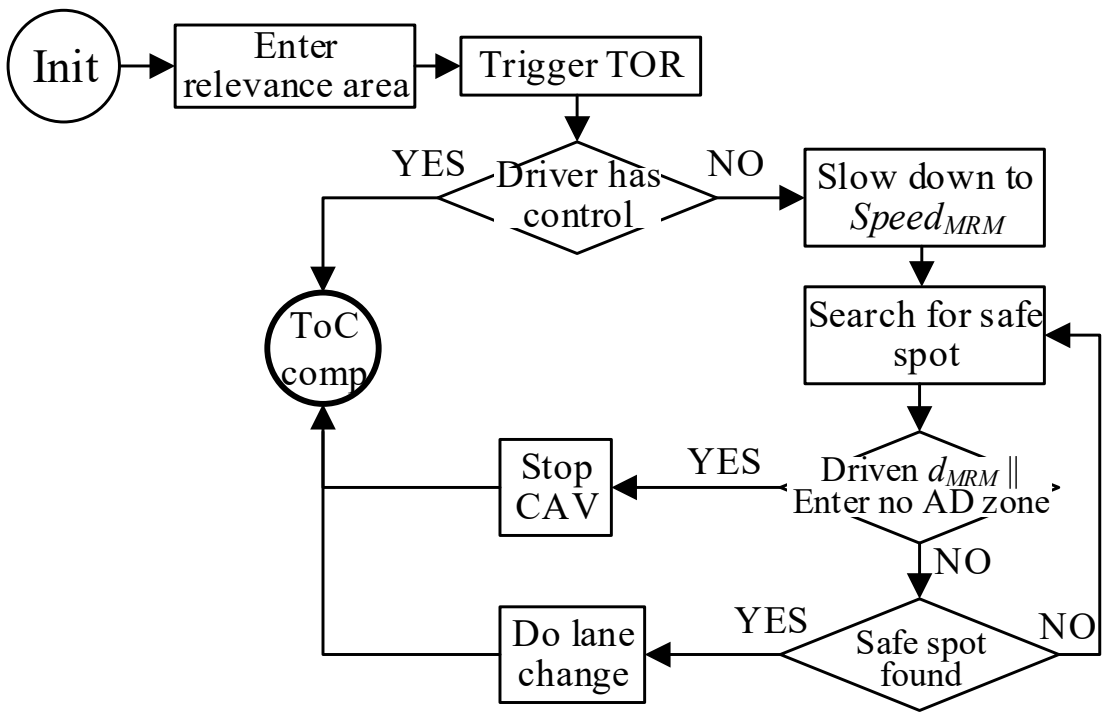

Fig. 12. DENM-based ToC processing at the AD SW.

### *B. MCM-based ToC management*

The RSU transmits MCM messages that include *TransitionOfControl* and *SafeSpot* advices to the vehicles approaching the no AD zone. When the CAV receives an MCM from the RSU, its AD SW analyzes the content received through the *IF1_MCM* (see Fig. 9) to understand how the RSU is suggesting the CAV to perform the ToC-related actions. Fig. 13 shows the ToC processing at the AD SW for the MCM-based ToC scheme. Using the *TransitionOfControl* and *SafeSpot* advices, the CAV identifies when/where it should request the driver to take over control (i.e., execute the TOR) and the location of a safe spot in case of MRM. We implement four different variants of the MCM-based ToC management scheme based on decisions that the RSU and the CAV could make to deal safely with the ToC.

The two following options are implemented at the RSU to suggest the CAV when it should issue the TOR:

- $min_{dMRM}$. The RSU schedules the execution of the TOR at the CAV so that it reaches the assigned safe spot driving the minimum possible distance at $Speed_{MRM}$. To implement this option, the RSU first selects a safe spot for the CAV, and takes into account its current location and driving speed (available through the CAMs) to identify the location where the TOR should be issued. The RSU takes a conservative decision on the selection of the location where the TOR should be executed since it is not feasible to assume it has a perfect knowledge of the CAV's implementation and dynamics (e.g., its deceleration).
- *DistrToC.* The RSU schedules the execution of the ToCs to distribute spatially their locations. To this aim, the RSU distributes the locations where to start the execution of the TORs within certain limits that are derived based on the information the RSU receives from the CAVs (e.g., the information included in the CAM and MCM messages). The

[2] The selected $t_{TOR}$ is within typical ranges [26]. However, it is out of the scope of this work to study the impact of $t_{TOR}$.

[3] This value has been selected after a series of field test runs. This value ensured the smoothest and more comfortable lane change and stop maneuver to the vehicle passengers.

[4] Processing a DENM only when the vehicle is in the relevance distance is in line with current implementations of hazard location warning applications. Relevance distances of 500m are typically used in current V2X infrastructure deployment scenarios like C-Roads: https://www.c-roads.eu/platform.html.

RSU also selects the location of the ToC with the restriction that the CAVs reach the assigned safe spot at $Speed_{MRM}$. The location of the ToC is therefore delimited by the current location of the CAV and the ToC location derived by $min_{dMRM}$ that minimizes the distance traveled at $Speed_{MRM}$. Then, and without loss of generality, this study has considered that the RSU randomly selects the locations where the CAVs issue the TOR within these limits.

At the CAV, we implement the following two options regarding the decisions the CAV could take after the TOR's lead time expires (i.e., during the execution of the MRM):

- *RSUadvice* (*rsu* for short). The CAV follows the RSU advices and slows down to $Speed_{MRM}$ as soon as the TOR's lead time expires. The CAV then drives at $Speed_{MRM}$ towards the assigned safe spot.
- *CAVdecision* (*cav* for short). The CAV uses the advice provided by the RSU and the knowledge of its own dynamics to decide the location at which it reduces the speed to $Speed_{MRM}$. In this study, the CAV determines that it is safer to maintain its driving speed when the TOR's lead time expires and slow down to $Speed_{MRM}$ only just before reaching the assigned safe spot to execute the parking maneuver[5].

The four variants of the MCM-based ToC management scheme combine the implementation options at the RSU and CAV and are referred to as: MCM_$min_{dMRM}$_*rsu*, MCM_$min_{dMRM}$_*cav*, MCM_*DistrToC_rsu*, MCM_*DistrToC_cav*.

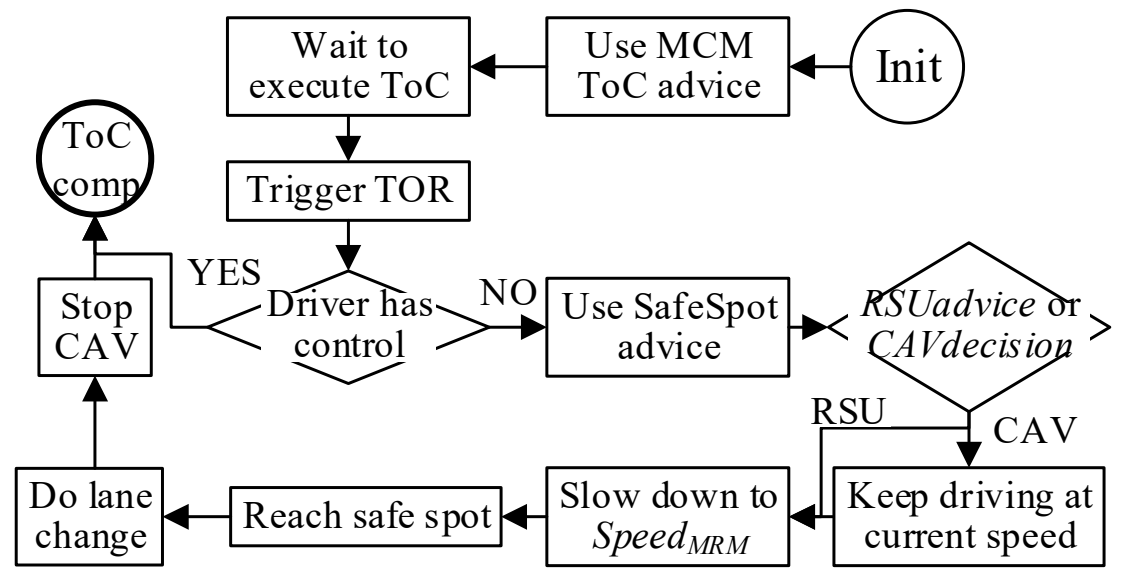

Fig. 13. MCM-based ToC processing at the AD software.

## VI. Results

We compare the performance of the DENM-based and MCM-based ToC management schemes using the following safety and traffic efficiency key performance indicators (KPIs):

- *Successful MRM*: represents the percentage of times the CAV executes a safe MRM. A safe MRM is achieved when the CAV is able to stop at a safe spot.
- *Distance to no AD Zone*: measures the distance to the no AD zone at which the CAV stops when it fails executing a safe MRM. In this case, the vehicle stops on the driving lane.
- *Distance traveled at* $Speed_{MRM}$: measures the distance that the CAV drives at speed $Speed_{MRM}$.
- *Distribution of ToC*: shows the Probability Distribution Function (PDF) of the spatial distribution of the performed ToCs.

The empirical results reported in the following sub-sections are average values of these KPIs measured over 50 field tests. Tests are conducted under 8 different scenario configurations that change the location of the safe spots (at least a safe spot is available in each scenario). A video that shows the execution of the MCM-based ToC management scheme is available at this link: https://www.transaid.eu/videos/. We complement the field trials with numerical evaluations in MATLAB that cover all possible combinations of scenarios (e.g., locations of safe spots) and configurations of the evaluated infrastructure-assisted ToC management schemes. This allows verifying the trends observed in the field trials over a larger number of tests and scenarios. To this aim, we utilize the conducted field tests to characterize the behavior of the CAV during the execution of the ToC when implementing the DENM- and MCM-based ToC management schemes. This characterization includes deriving the distances traveled by the CAV for the execution of each of the actions needed to complete the ToC. These distances are represented in Table I.

TABLE I
Distances traveled by the CAV for the execution of the ToC.

| Variable | Value | Meaning |
|---|---|---|
| $d_{TOR}$ | 166m | Distance traveled during the execution of the TOR when the vehicle drives at 60 Km/h for 10s |
| $d_{2SpeedMRM}$ | 150m | Distance traveled for decelerating to $Speed_{MRM}$ (i.e., from 60 Km/h to 20 Km/h) |
| $d_{2Stop}$ | 24m | Distance traveled to stop the vehicle from $Speed_{MRM}$ |
| $d_{LC}$ | 68m | Distance traveled to change from the driving lane to the emergency lane |

Using this characterization, we numerically resolve the flow charts in Fig. 12 and Fig. 13 to evaluate the performance of the DENM- and MCM-based ToC management schemes under the scenarios and conditions that could not be tested over the field trials. We also randomly select the safe spots in the numerical evaluation. The DENM-based ToC management scheme requests CAVs issuing the TOR at *relevanceDistanceDENM* meters to the no AD zone. The *relevanceDistanceDENM* is set to 500m. The CAV reaches $Speed_{MRM}$ when it is at *relevance DistanceDENM* – ($d_{TOR}$ + $d_{2SpeedMRM}$) meters to the no AD zone. The CAV performs a successful MRM if the range of locations [*relevanceDistanceDENM* – ($d_{TOR}$ + $d_{2SpeedMRM}$), *relevance DistanceDENM* – ($d_{TOR}$ + $d_{2SpeedMRM}$ + $d_{MRM}$)] intersect the location of an available safe spot. The proper execution of the MRM also needs that the location of the safe spot that first reaches the CAV at $Speed_{MRM}$ falls $d_{LC}$ meters away to the end of the safe spot. Otherwise, the CAV would drive to the location *relevanceDistanceDENM* – ($d_{TOR}$ + $d_{2SpeedMRM}$ + $d_{MRM}$ + $d_{2Stop}$) and would stop on the driving lane.

The numerical evaluation of the MCM-based ToC management scheme considers the two different implementation options at the RSU to schedule the ToC. For the *DistrToC* option, the location to issue the TOR is randomly selected in the range [*Max_ToCrange*, *Min_dist2SafeSpot*] following an uniform distribution. *Max_ToCrange* represents the furthest location from the no AD zone where the CAV can receive the MCM from the RSU. In the field trials, this location is 700m and it is imposed by the size of the scenario. The numerical evaluation has extended it to 900m in line with V2X coverage ranges[6]. *Min_dist2SafeSpot* represents the closest location to an available safe spot where the CAV can issue the TOR. *Min_dist2SafeSpot* is determined in the numerical evaluation taking into account the location of an available safe spot and the distance it takes to the CAV to reach the safe spot at $Speed_{MRM}$, i.e., $d_{TOR} + d_{2SpeedMRM} + Y$. $Y$ is the conservative margin considered at the RSU because of the unknown deceleration of the CAV, and it is set to 15m. For the $min_{dMRM}$ option, the location derived to perform the ToC is *Min_dist2SafeSpot*. The CAV keeps its driving speed from *Max_ToCrange* until it reaches the derived TOR location. For the *RSUadvice* option, the CAV slows down to $Speed_{MRM}$ when the TOR's lead time expires. For the *CAVdecision* option, the CAV keeps its driving speed until it is $d_{2SpeedMRM}$ apart from the assigned safe spot, and then it slows down to $Speed_{MRM}$.

[5] An equivalent variant for the DENM-based ToC approach is not feasible. This is the case because the DENM does not include information about the available safe spots and the CAV has to discover them using its own sensors while driving. When it finds one, the CAV has to slow down to $Speed_{MRM}$ before executing the parking maneuver. It would pass the safe spot if it is still driving at its cruise speed.

### A. Successful MRM

Table II.a reports the rate of successful MRMs with the two infrastructure-assisted ToC management schemes during the field trials. The empirical results show that when the CAV follows the DENM-based ToC management scheme, it does not always successfully implement a safe MRM. It is important to recall that the DENM's relevant information is only made available at the AD SW once the CAV is within the relevance distance (i.e., 500m away to the no AD zone). At this point in time, the AD SW triggers the TOR, and the CAV slows down from its driving speed to $Speed_{MRM}$. Therefore, the CAV misses any safe spot available from the start of the DENM's relevance area to the point at which it reaches $Speed_{MRM}$. Table II.a shows that when the CAV uses the DENM-based ToC management scheme with $d_{MRM}$=0, it only finds 12.5% of the times a safe spot available where to perform the parking maneuver. In this case, the CAV stops on the driving lane in 87.5% of the tests. Table II.a shows that the DENM-based ToC management scheme benefits from a higher $d_{MRM}$. The CAV performs a successful MRM 50% and 62.5% of the times when $d_{MRM}$ is set to 50m and unlimited, respectively. On the other hand, Table II.a shows that all variants of the MCM-based ToC management scheme always perform a successful MRM. The MCM-based ToC management scheme benefits from the *TransitionOfControl* and *SafeSpot* advices transmitted by the RSU that inform the CAV when/where it should execute the ToC so that it can reach the assigned safe spot in case of MRM.

Tables II.b and II.c report average values for the successful MRM KPI obtained numerically. The results reported in Table II.b correspond to a scenario where there is one safe spot available on the emergency lane to park the CAV. Table II.c reports results when there are two safe spots. In both scenarios, the safe spots are randomly chosen as it is the case during the field trials. The numerical evaluation covers all possible combinations of locations of safe spots. The numerical results reported in Tables II.b and II.c show similar trends to those analyzed in Table II.a for the field trials. CAVs implementing the DENM-based ToC management scheme benefit from a higher $d_{MRM}$ to successfully park on a safe spot during the MRM. In addition, Table II.c shows that CAVs implementing the DENM-based ToC management scheme are more likely to find a safe spot with the increasing number of available safe spots in the scenario. However, the DENM-based ToC management scheme cannot always perform a successful MRM as it is the case of the MCM-based ToC management scheme.

TABLE II
SUCCESSFUL MRM.

a) Field trials (at least 1 safe spot available)

| **DENM_$d_{MRM}$** | | | **MCM_$min_{dMRM}$** | | **MCM_*DistrToC*** | |
|---|---|---|---|---|---|---|
| **0** | **50** | ***unlimited*** | ***rsu*** | ***cav*** | ***rsu*** | ***Cav*** |
| 12.5% | 50% | 62.5% | 100% | 100% | 100% | 100% |

b) Numerical evaluation (1 safe spot available)

| **DENM_$d_{MRM}$** | | | **MCM_$min_{dMRM}$** | | **MCM_*DistrToC*** | |
|---|---|---|---|---|---|---|
| **0** | **50** | ***unlimited*** | ***rsu*** | ***cav*** | ***rsu*** | ***Cav*** |
| 5.5% | 16.5% | 33.5% | 100% | 100% | 100% | 100% |

c) Numerical evaluation (2 safe spots available)

| **DENM_$d_{MRM}$** | | | **MCM_$min_{dMRM}$** | | **MCM_*DistrToC*** | |
|---|---|---|---|---|---|---|
| **0** | **50** | ***unlimited*** | ***rsu*** | ***cav*** | ***rsu*** | ***Cav*** |
| 13% | 33% | 62.3% | 100% | 100% | 100% | 100% |

### B. Distance to no AD zone

Table II shows that CAVs implementing the DENM-based ToC management scheme cannot always find a safe spot. When this happens, the CAVs perform the MRM by stopping on the driving lane. A vehicle blocking a free lane might reduce the traffic flow capacity and could cause safety risks. The situation might get worse the closer to the no AD zone the CAV stops. This is the case because of the increasing risk of completely blocking the traffic coming from behind. For the conducted field trials, Table III.a shows that this situation is more likely to happen when the DENM-based ToC management scheme is configured with a higher $d_{MRM}$. For example, when the DENM-based ToC management scheme is configured with $d_{MRM}$=0, the CAV stops ~160m away to the no AD zone. This distance reduces as $d_{MRM}$ increases. In this case, the CAV can drive for longer distances while in MRM and this increases the likelihood to find a safe spot (Table II). However, the CAV does not have enough information that guarantees the availability of such safe spot. This can result in that the CAV stops just in front the start of the no AD zone. Table III.a shows that this is actually the case for the DENM-based ToC management scheme when it is configured with $d_{MRM}$=unlimited. It should be noted that empirical results are not reported in Table III.a for the MCM-based ToC management scheme since all MRM maneuvers were safely conducted and hence the CAV never stopped on the driving lane (Table II).

[6] Note that the larger value of *Max_ToCrange* in the numeral evaluation compared to the field trials (i.e., 900m vs 700m) results in an increased range where TOR can be issued when the RSU implements the *DistrToC* option. On the other hand, the *relevanceDistanceDENM* is set to 500m. Moving away it from the no AD zone does not have any impact on the distribution of the ToCs because they are all performed when CAVs reach the *relevanceDistanceDENM*.

The obtained numerical results are reported in Table III.b. Results are independent on the number of safe spots available, since the distance to the no AD zone is actually measured when the CAV cannot stop in a safe spot. In this case, the distance to no AD zone is calculated using the derived distances from the conducted field tests as: *relevanceDistanceDENM* – ($d_{TOR}$ + $d_{2SpeedMRM}$ + $d_{MRM}$ + $d_{2Stop}$). Therefore, the numerical results coincide closely with the empirical results. The small differences are due to the fact that the numerical evaluation does not model all external conditions that slightly alter the empirical results, e.g., that the CAV does not drive completely straight.

TABLE III
DISTANCE TO THE NO AD ZONE WHEN CAV STOPS ON THE DRIVING LANE.
a) Field trials (at least 1 safe spot available)

| DENM_ $d_{MRM}$=0 | DENM_ $d_{MRM}$=50 | DENM_ $d_{MRM}$=unlimited |
|---|---|---|
| 161.1m | 109.8m | 2.3m |

b) Numerical evaluation (1 & 2 safe spots available)

| DENM_ $d_{MRM}$=0 | DENM_ $d_{MRM}$=50 | DENM_ $d_{MRM}$=unlimited |
|---|---|---|
| 160m | 110m | 0m |

## C. Distance traveled at MRM speed

A slow-moving vehicle can negatively impact the traffic flow and, depending on the scenario, generate a safety risk. Fig. 14.a depicts the distance that the CAV travels at speed $Speed_{MRM}$ during the field trials. On each box, the central red horizontal mark is the median, the edges of the box are the 25th and 75th percentiles, and the whiskers extend to the 5th and 95th percentiles. The DENM-based ToC management scheme uses the $d_{MRM}$ to limit the distance that the CAV can travel at speed $Speed_{MRM}$ while searching for an available safe spot. Fig. 14.a shows that the DENM_$d_{MRM}$=50 and DENM_$d_{MRM}$=unlimited variants limit the distance traveled at speed $Speed_{MRM}$ to 50m and 160m, respectively[7]. The CAV might find an available safe spot before reaching this limit. This is the reason why the median value of the measurements for the distance traveled at speed $Speed_{MRM}$ is 45m and 65m, respectively. The 25th percentile is below 15m for both cases.

The variants of the MCM-based ToC management scheme that implement at the CAV the *RSUadvice* option (i.e., MCM_$min_{dMRM}$_rsu and MCM_*DistrToC_rsu*) are also configured to slow down to speed $Speed_{MRM}$ when the MRM starts. For the MCM_$min_{dMRM}$_rsu variant, the CAV would ideally reach the speed $Speed_{MRM}$ just before the assigned safe spot. However, this would be achieved if the RSU had a perfect knowledge on the CAV's dynamics, including its deceleration. However, this might not be the case since the RSU is only aware of the information included in the CAM and MCM messages. The RSU was then configured to issue advices with conservative decisions in the scheduling of the ToC that assume a slower deceleration of the CAV. This results in that the CAV reaches the speed $Speed_{MRM}$ some meters before the start of the assigned safe spot. The measurements reported in Fig. 14.a show that, in the worst case, these conservative decisions at the RSU resulted in that the CAV drove ~70m at speed $Speed_{MRM}$. The results reported in Fig. 14.a for the *MCM_DistrToC_rsu* variant show that this distance can be even longer when the RSU follows the *DistrToC* option. In this case, the CAV drove up to 280 meters at speed $Speed_{MRM}$ before reaching the safe spot. The MCM_$min_{dMRM}$_cav and MCM_*DistrToC_cav* variants are based on the same exact ToC management options at the RSU. Then, they issued the TOR at the same exact locations. However, MCM_$min_{dMRM}$_cav and *MCM_DistrToC_cav* implement at the CAV the *CAVdecision* option. Fig. 14.a shows that this allows the CAV to reach the safe spot without having to drive at speed $Speed_{MRM}$. Under the *CAVdecision* option, the CAV checks its distance to the assigned safe spot when the MRM starts. Then, it calculates the distance it needs to decelerate from its current speed to the $Speed_{MRM}$. Contrary to the RSU, this information is available at the CAV. Then, the CAV continues driving at its current speed and only slows down to $Speed_{MRM}$ when it identifies that it will reach $Speed_{MRM}$ just before the allocated safe spot to execute the parking maneuver.

Fig. 14.b shows very similar trends in the numerical evaluation to those analyzed in Fig. 14.a. The numerical evaluation has been executed for all different configurations in the scenario including the location of the safe spots. In addition, for the *DistrToC* option at the RSU, it allows considering all different locations where the TORs are randomly issued. In this context, the wider set of experiments conducted through the numerical evaluation are at the origin of the differences with the empirical results when, for example, the MCM_*DistrToC_rsu* variant is compared. On the other hand, the exact reproducibility of the experiments in the numerical evaluation results in less spread outputs than in the real-world experiments. This explains why the CAV implementing the MCM_$min_{dMRM}$_rsu variant always reaches the $Speed_{MRM}$ 49m before the assigned safe spot. This value matches with the median of the measurements in the empirical results.

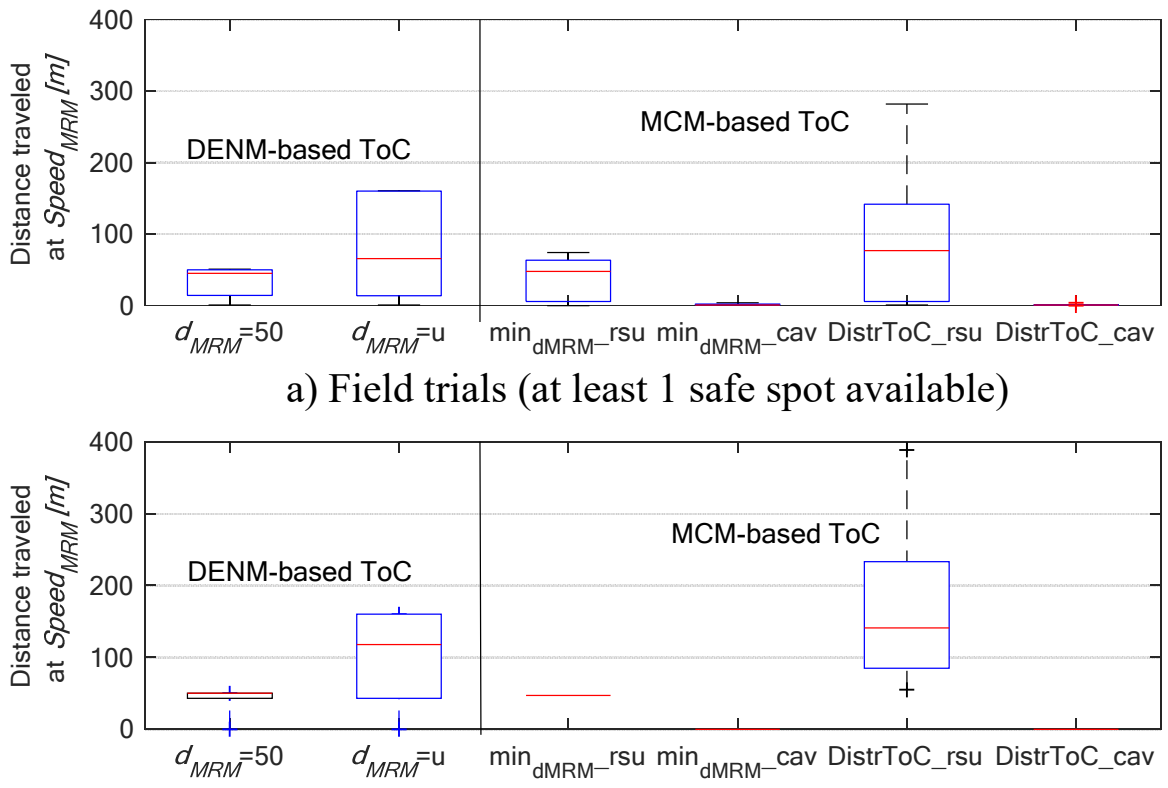

a) Field trials (at least 1 safe spot available)

b) Numerical evaluation (2 safe spots available)[8]

Fig. 14. Distance traveled at speed $Speed_{MRM}$.

## D. Distribution of ToC

The study in [8] showed through simulations that the impact of the ToC management on the traffic safety and efficiency depends on how distributed are the ToCs when CAVs head towards a no AD zone. In the case of the DENM-based ToC, CAVs perform the ToC by requesting the driver to take over control as soon as they enter the DENM's relevance area. Therefore, CAVs approaching the no AD zone perform the ToC at the exact same location $x$ that is *relevanceDistanceDENM* meters away to the no AD zone. The PDF ($f$) of the distribution of the DENM-based ToC management scheme as a function of the distance to the no AD zone can then be expressed as:

$$f_{DENM}(x)=\delta(x - relevanceDistanceDENM), \quad (1)$$

where $\delta(x)$ is a Dirac delta function that is equal to 1 when $x$=0, and 0 otherwise. Fig. 15 shows the empirical (Fig. 15.a) and numerical (Fig. 15.b) distribution of the ToCs for the DENM-based ToC management scheme as a function of the distance of the CAV to the no AD zone. As derived from (1), the results show that ToCs are always performed at the location 500m that coincides with the configured DENM's relevance distance.

[7] DENM_ $d_{MRM}$=0 has not been included in the results reported in Fig. 14 as this variant does not allow the CAV travelling at MRM speed.

[8] Similar trends were observed for the scenario with 1 safe spot.

The MCM-based ToC management scheme that implements at the RSU the $min_{dMRM}$ option seeks minimizing the distance that the CAVs travel at $Speed_{MRM}$. The $min_{dMRM}$ option links each possible safe spot with a location where to perform the ToC. Considering that all potential safe spots are independent and equally usable, the PDF of the distribution of the ToCs for the $min_{dMRM}$ option at the RSU can be expressed as:

$$f_{mindMRM}(x)= \frac{1}{n}\sum_{i=0}^{n-1} \delta\left(x - i{\cdot}S_{len} - d_{ToC} - SafeSpot_{len}\right). \quad (2)$$

In (2), $S_{len}$ is the length of the sections on the emergency lane (i.e., 25m for this study), and $n$ is the number of safe spots that can be found in the emergency lane. $d_{ToC}$ is the sum of the distances traveled by the CAV during the TOR's lead time and deceleration from driving speed to $Speed_{MRM}$. $SafeSpot_{len}$ is the length of a safe spot (i.e., 75m since it is made of 3 25-m segments). The expression derived in (2) shows that with the $min_{dMRM}$ option at the RSU, the ToCs are performed at discrete and equally spaced locations that are linked to the $n$ safe spots. Both the empirical and the numerical evaluations were executed with the assumption that the safe spots are only available within the DENM's relevance area. Therefore, there are $n$=18 potential safe spots. The results reported in Fig. 15.b for the $min_{dMRM}$ option show these 18 locations where the ToCs are performed. In addition, the empirical measurements show that $d_{ToC}$ is approximately equal to 325m (i.e., ~166m driven at 60 Km/h during the 10s TOR's lead time + ~150 m for decelerating from 60 Km/h to 20 Km/h + conservative margin at the RSU). Then, based on (2), the nearest location to the no AD zone where ToCs are performed is at 400m to the no AD zone. From this location, ToCs' triggering locations are 25m equally spaced and probable. Fig. 15.a shows that the results obtained for the MCM_$min_{dMRM}$ variant in the numerical evaluation follow the analytical formula derived in (2). Field tests were not conducted for all possible locations where the safe spots could be. For the conducted field tests, the results reported in Fig. 15.a for the MCM_$min_{dMRM}$ variant also correlates well with the analytical expression in (2) and the numerical results in Fig. 15.b.

The MCM-based ToC management scheme that implements at the RSU the *DistrToC* option seeks maximizing the spatial distribution of the ToCs. As detailed in Section V.B, the location where the ToC is performed is randomly chosen by the RSU between the current location of the CAV and the location derived following the $min_{dMRM}$ option (this depends on the location of the safe spot). To compute the PDF of the distribution of the ToC for the MCM_*DistrToC* variant, we use (2) to derive the locations $x$ where the ToC can be performed in order to reach the safe spot that is located just before the no AD zone. Locations $x$ must satisfy:

$$d_{ToC} + SafeSpot_{len} + S_{len} > x \geq d_{ToC} + SafeSpot_{len}. \quad (3)$$

The probability to perform a ToC in these locations is:

$$P_1(x)=P_{park}/ToC_{range}. \quad (4)$$

In (4), $P_{park}$=1/$n$ is the probability that this safe spot is free, and $ToC_{range}$ is the length of the range of locations where ToCs can be performed. This range is limited by the furthest and closest distance to the no AD zone where a ToC can be performed. To reach the two closest safe spots to the no AD zone, the ToC could be performed at any location $x$ that satisfies:

$$d_{ToC}+SafeSpot_{len}+2S_{len}> x\geq d_{ToC}+SafeSpot_{len}+S_{len}. \quad (5)$$

In this case, the probability that each of these locations are selected to perform the ToC is

$$P_2(x)= P_1(x)+P_{park}/(ToC_{range} - S_{len}). \quad (6)$$

Note that the first term of $P_2(x)$ corresponds to the case in which the safe spot that is available is the one closest to the no AD zone ('1'), and the second term when the available safe spot is the next one ('2'). Following this procedure, to reach the safe spots '1', ..., '$k$' | $k < n$, the ToC could be performed at any location $x$ that satisfies:

$$d_{ToC} + SafeSpot_{len}+k{\cdot}S_{len} > x\geq d_{ToC}+SafeSpot_{len}+(k-1)S_{len}. \quad (7)$$

The probability that each of the locations in (7) are selected can be computed as:

$$P_k(x)= P_{k-1} + P_{park}/(ToC_{range} - (k-1)S_{length}),\ \forall k \mid k < n. \quad (8)$$

Finally, to reach any of the safe spots, the ToC could be performed at any location $x$ that satisfies:

$$Max_ToCrange > x \geq d_{ToC}+ (n-1){\cdot}SafeSpot_{len}, \quad (9)$$

where *Max_ToCrange* is the furthest distance to the no AD zone where a ToC can be performed. The probability that each of the locations in (9) are selected is computed as:

$$P_n(x)= P_{n-1} + P_{park}/(ToC_{range} - (n-1){\cdot}S_{len}). \quad (10)$$

Therefore, the PDF of the distribution of the ToC for the MCM_*DistrToC* variant can be computed using (3)-(10) as:

$$f_{DistrToC}(x)= \{P_1(x), ..., P_n(x)\}. \quad (11)$$

Fig. 15.b shows the spatial distribution of the ToCs for the MCM_*DistrToC* variant obtained in the numerical evaluation and that follows the analytical expression derived in equation (11). Fig. 15.b demonstrates how the MCM_*DistrToC* variant achieves a higher spatial distribution of the ToCs than the MCM_$min_{dMRM}$ variant. The numerical results complement the empirical results reported in Fig. 15.a that also show that the ToCs for the MCM_*DistrToC* variant are not performed at discrete locations linked to the safe spots, and that they are more likely to be performed at longer distances to the no AD zone. Due to space limitations in the proving ground of the Griesheim airport (Fig. 11), ToCs could not be performed beyond 700m to the no AD zone during the field trials. The numerical results reported in Fig. 15.b have been derived covering all possible random locations where ToCs would be performed in the scenario (this was not feasible in field tests). For the set of conducted field tests, the empirical results reported in Fig. 15.a show a good correlation with the numerical results and confirm the spatial distribution of ToCs that can be achieved with the MCM_*DistrToC* variant.

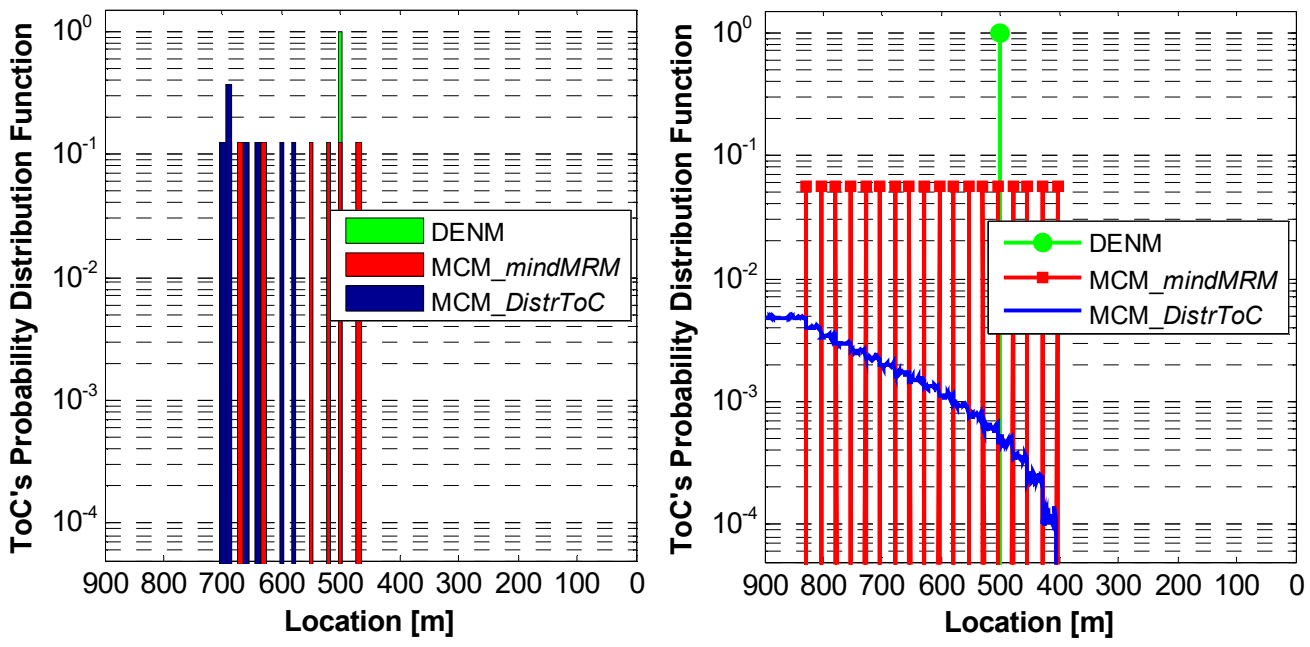

a) Field trials b) Numerical evaluation

Fig. 15. Distribution of ToCs.

## VII. Discussion and Conclusions

This paper has presented the design, implementation and usage of a prototype platform for validation and testing of infrastructure-assisted ToC management for CAVs. The platform includes an RSU and a ROS2-programmed CAV that have been equipped with a V2X module to enable the implementation of infrastructure-assisted cooperative traffic management measures. These measures are designed to manage ToCs and MRMs with the objective to mitigate the potential negative effects of ToCs in Transition Areas. If a driver does not respond to a ToC request, the implemented platform allows the CAV to search for a safe spot as part of an MRM maneuver. The study has compared the performance of two ToC management schemes that rely on standard V2X messages. First, a DENM-based ToC management scheme is utilized by the RSU to inform the CAV about the presence of a no AD zone before reaching it. The CAV triggers a ToC as soon as it enters the DENM's relevance area. The conducted field tests have shown that this DENM-based scheme can result in that all CAVs perform the ToC at the exact same location. This can cause traffic disruption and safety risks. In addition, the DENM does not provide information about available safe spots to park. This can lead to a CAV stopping on the driving lane if it is not capable to find a safe spot during an MRM. The conducted trials showed that augmenting the distance over which a CAV can search for a safe spot increases the likelihood of finding one. However, this is at the expense of driving at low speeds for longer distances and this entails safety risks and traffic disruptions to nearby vehicles.

This study has also implemented and evaluated an MCM-based ToC management scheme. The use of MCM messages provides additional information that the infrastructure or RSU can use to give individual advices to vehicles and provide specific information about safe spots to facilitate MRMs. Different variants of the MCM-based ToC management scheme have been implemented. These variants can modify the configuration of ToC and safe spot advices based on the sought objective. For example, the RSU can opt for scheduling ToCs in order to minimize the distance CAVs drive at low speed in case the driver does not take control and an MRM is performed. This strategy can result in that close-driving CAVs that are assigned the same safe spot are also advised to execute a ToC at the same point. Another variant is implemented for the RSU to distribute ToC locations at distinct CAVs with the objective to minimize their impact on the surrounding traffic. The conducted tests have shown that this scheme can address some of the previous challenges but can also make CAVs travel long distances at low speed to reach the assigned safe spot in case of MRM. This study has also proposed different variants to address these inefficiencies. The conducted tests have also shown that providing the possibility for the RSU to give individual advices to CAVs about safe spots using MCM messages allows for a safer execution of MRMs. In particular, the implemented CAV can be configured to continue driving at its cruise speed while in MRM and slow down only in proximity of the assigned safe spot. The authors acknowledge that this measure might not be exempt of risk, since this is achieved by allowing the CAV to drive at its current speed during MRM. Nevertheless, it has shown to minimize the distance that the CAV is driving at low speed and hence might imply lower risks for the surrounding traffic.

In summary, the conducted field tests have demonstrated that the infrastructure and V2X technologies can play an important role to implement cooperative traffic management measures that facilitate the introduction of CAVs and improve the execution of ToCs and MRMs. The infrastructure can provide individual advices to vehicles to effectively prevent multiple CAVs from executing ToCs at close by locations, hence diminishing potential safety risks deriving from such situations. These measures also support CAVs much better in finding a safe spot for parking and therefore improve the safety of MRMs. To this aim, the infrastructure can exploit cooperative maneuvering and MCM messages. The use of MCM messages and an adequate configuration of the infrastructure-assisted ToC management scheme can reduce or eliminate the probability for CAVs to block or stop on driving lanes. Stopping a CAV on the driving lane can block traffic and generate significant traffic risks. The combined use of information available at the CAV with that received from the infrastructure can help CAVs prevent from having to drive at low speeds which can negatively impact the traffic flow. In this study, the combined information from the infrastructure and the CAVs has been utilized to reach the safe spots in case of MRMs at cruise speed, which helps reduce or eliminate traffic disruptions or safety risks.

In this context, the main findings of this study are as follows:

- A characterization of the CAV during the execution of the DENM- and MCM-based ToC management schemes is derived that reports the distances traveled by the CAV while implementing the actions needed to complete to ToC.
- The conducted experiments show that the MCM-based ToC management scheme always succeeds in guiding CAVs to a safe spot where to park in case of MRM.
- The success rate of CAVs that find a safe spot in MRM when they implement the DENM-based ToC management

scheme is much lower (~13%) than the one based on MCM messages. This rate increases at the expense of potential safety risks like having stopped vehicles closer to the no AD zone and allowing vehicles to drive longer distances at low MRM speeds.

- The MCM-based ToC management scheme can empower CAVs to minimize the distance they travel at MRM speed when they combine the advices from the infrastructure to execute the ToC and their own information. Under the considered scenario, CAVs that just follow the infrastructure advices (without additionally using local knowledge) could drive up to 280m at MRM speeds in the worst case before reaching the safe spot.
- The DENM-based ToC management scheme does not distribute ToCs in time or in space since it requests the driver to take over control as soon as they enter the DENM's relevance area. This is to be avoided as the execution of multiple ToCs at close location can turn into risky situations.
- Implementing the MCM-based ToC management schemes can help CAVs distribute their ToCs. If the infrastructure schedules the ToCs based on the available safe spots, CAVs perform the ToCs at discrete locations that are linked to the safe spots, obtaining a sub-optimal distribution. The infrastructure can also schedule ToCs to maximize their spatial distribution while guaranteeing the CAVs reach an available safe spot.

The proposed infrastructure-assisted ToC management scheme is the first proposal that exploits MCM messages to give individual advices to CAVs with the aim to mitigate the potential negative effects of ToCs in Transition Areas. An extension of this scheme could focus on the designing efficient ToC management schemes at the infrastructure that simultaneously take into account advices already being executed by multiple CAVs and advises that are still scheduled. In this context, the ToC management scheme could also take into account, as a feedback loop variable, notifications of whether a given advice is currently being followed by a CAV (this capability is also supported by the proposed MCM extensions). Extensions of this work could also focus on comparing the effectiveness of infrastructure-assisted and distributed (e.g., based on V2V, or non-communicating AVs) traffic management approaches. Distributed solutions might be still needed when, e.g., the infrastructure is not available. The design of efficient distributed solutions might be challenging, especially in areas where multiple ToC and maneuvers need to be simultaneously coordinated.

## References

[1] L3Pilot Consortium, "Testing of Automated Driving on Public Roads: Challenges and First Lessons Learned", *26th ITS World Congress*, 21-25 October, Singapore, 2019. [Online]: https://l3pilot.eu/download/
[2] TransAID Consortium, "Use cases and safety and efficiency metrics", Deliverable D2.1, 2018.
[3] A. Eriksson and N. A. Stanton, "Driving Performance after Self-Regulated Control Transitions in Highly Automated Vehicles", *Human Factors, vol. 59, pp. 1233-1248*, 2017. DOI: 10.1177/0018720817728774.
[4] H.Clark and J. Feng, "Age differences in the takeover of vehicle control and engagement in non-driving-related activities in simulated driving with conditional automation", *Accident Analysis & Prevention, vol. 106, pp. 468–479*, 2017. DOI: 10.1016/j.aap.2016.08.027.
[5] B. Mok, et al., "Tunneled In: Drivers with Active Secondary Tasks Need More Time to Transition from Automation", in *Proc. CHI*, pp. 2840–2844, New York, USA, 2017. DOI: 10.1145/3025453.3025713
[6] M.S. Young, and N.A. Stanton, "Back to the future: Brake reaction times for manual and automated vehicles", *Ergonomics, vol. 50, no. 1, pp. 46–58*, 2015. DOI: 10.1080/00140130600980789
[7] HAVEit consortium , "Function description and requirements", Deliverable D11.1, 2008.
[8] TransAID Consortium, "Preliminary simulation and assessment of enhanced traffic management measures", Deliverable D4.2, 2019.
[9] TransAID Consortium, "Modelling, simulation and assessment of vehicle automations and automated vehicles' driver behaviour in mixed traffic", Deliverable D3.1, 2019.
[10] L. Lücken, E. Mintsis, K. N. Porfyri, R. Alms, Y.-P. Flötteröd, and D. Koutras, "From Automated to Manual-Modeling Control Transitions with SUMO", *SUMO 2019, EPiC Series in Computing, Vol. 62, pp 124-144*, Berlin, Germany, 2019.
[11] A. Morando, et al., "Users' Response to Critical Situations in Automated Driving: Rear-Ends, Sideswipes, and False Warnings", *IEEE Transactions on Intelligent Transportation Systems*. DOI: 10.1109/TITS.2020.2975429.
[12] N. Merat, et al., "Transition to manual: Driver behaviour when resuming control from a highly automated vehicle", *Transp Res Part F Traffic Psychol Behav, vol. 27, part B, pp. 274-282*, 2014. DOI: 10.1016/j.trf.2014.09.005
[13] E. Mintsis, et al., "Joint Development of Infrastructure-Assisted Traffic Management and Cooperative Driving around Work Zones", *23rd IEEE ITSC*, Rhodes, Greece, 2020.
[14] ERTRAC Working Group on Connectivity and Automated Driving, "Connected Automated Driving Roadmap", Whitepaper, March 2019. [Online]: https://connectedautomateddriving.eu/wp-content/uploads/2019/04/ERTRAC-CAD-Roadmap-03.04.2019-1.pdf
[15] F. Sánchez, "CONCORDA project: Spanish Test Site", ITS European Congress, June 2019, Eindhoven, NL.
[16] ETSI, "ITS; Veh Comms; Basic Set of Apps; Part 3: Spec of Decentralized Environmental Notification Basic Service", EN 302 637-3, v1.2.2, 2014.
[17] ETSI, "Intelligent Transport Systems (ITS); Vehicular Communications; Informative Report for the Maneuver Coordination Service", TR 103 578, v0.0.5, May 2020.
[18] ETSI, "ITS; Veh Comms; Basic Set of Apps; Maneuver Coordination Service", TR 103 561, v0.0.1, 2018.
[19] A. Correa, et al., "Infrastructure Support for Cooperative Maneuvers in Connected and Automated Driving", *in Proc. IEEE IV*, Paris, France, 2019.
[20] TransAID Consortium, "System Prototype Demonstration", Deliverable D7.2, 2019.
[21] ETSI, "ITS; Veh Comms; Basic Set of Apps; Part 2: Specification of Cooperative Awareness Basic Service", EN 302 637-2, v1.4.1, 2019.
[22] SAE, "Taxonomy and Definitions for Terms Related to Driving Automation Systems for On-Road Motor Vehicles", J3016, 2018.
[23] MAVEN Consortium, "V2X communications for infrastructure-assisted automated driving", Deliverable 5.1, February 2018.
[24] IEEE Standard for Information technology, "LAN and MAN-Part 11: Wireless LAN MAC and PHY Spec Amendment 6: Wireless Access in Vehicular Environments", in IEEE Std 802.11p-2010, pp. 1-51, 15 2010.
[25] M. Reke, et al., "A Self-Driving Car Architecture in ROS2", in Proc. SAUPEC/RobMech/PRASA, pp. 1-6, Cape Town, South Africa, 2020.
[26] J. Wang and D. Söffker, "Bridging Gaps Among Human, Assisted, and Automated Driving with DVIs: A Conceptional Experimental Study", IEEE Transactions on Intelligent Transportation Systems, vol. 20, no. 6, pp. 2096-2108, June 2019. DOI: 10.1109/TITS.2018.2858179.

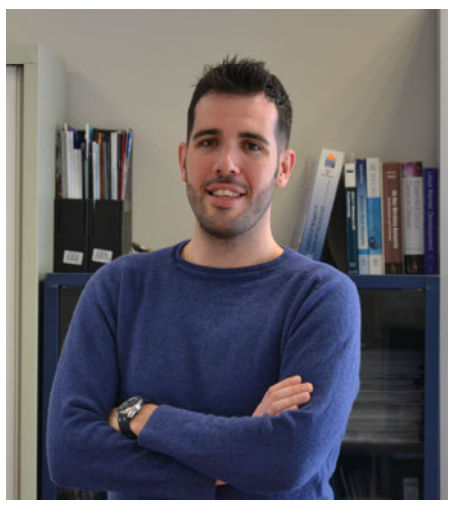

**Baldomero Coll-Perales** received the M.Sc.Eng. and Ph.D. degrees from the Universidad Miguel Hernandez (UMH) de Elche, Spain. He is currently a Research Fellow at the UWICORE laboratory. He was formerly Postdoctoral Associate at HMETC (Russelsheim, DE), IIT-CNR (Pisa, IT) and WINLAB (Rutgers University, NJ, USA). His research interests lie in the field of connected automated vehicles and device-centric technologies. He is Associate Editor for Springer's Telecommunication Systems and Int. J. of Sensor Networks. He has served as Track Co-Chair for IEEE VTC-Fall 2018, and as member of the TPC in over 30 international conferences.

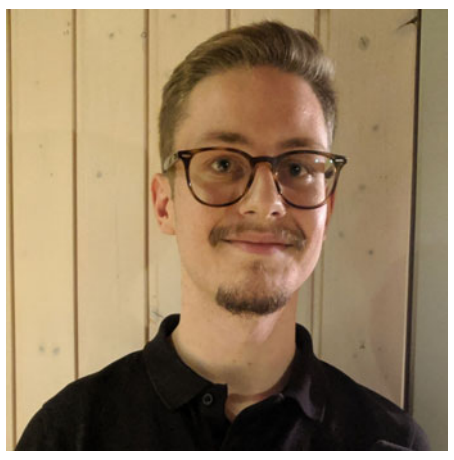

**Joschua Schulte-Tigges** holds a Computer Science degree from FH Aachen. He is pursuing his master's degree on Information System Engineering. He has been an intern with Hyundai Motor Europe (Germany) and is currently a researcher with FH Aachen working in the field of automated driving.

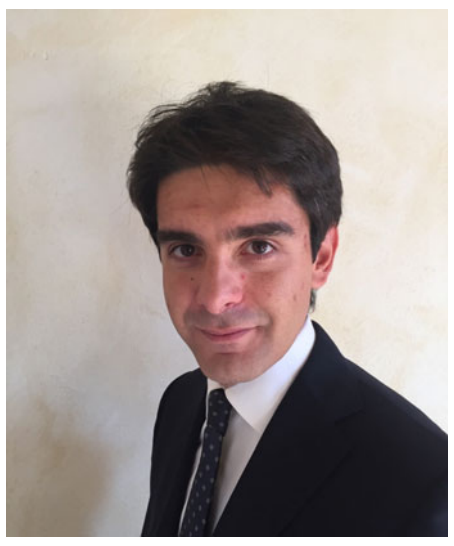

**Michele Rondinone** received a M.Sc. degree in Telecommunications Engineering from the University of Bologna (Italy), and a PhD in Industrial and Telecommunications Technologies from UMH. Since 2014 he is with the Hyundai Motor Europe Technical Center (Germany), where he works as Senior Engineer on V2X deployment preparation and future V2X for automated driving in internal as well as EU-funded projects. He actively participates in European V2X standardization at ETSI TC ITS and C2C-CC and served as TCP co-chair for the IEEE Connected and Automated Vehicles Symposium 2018 as well as TPC member for several IEEE international conferences.

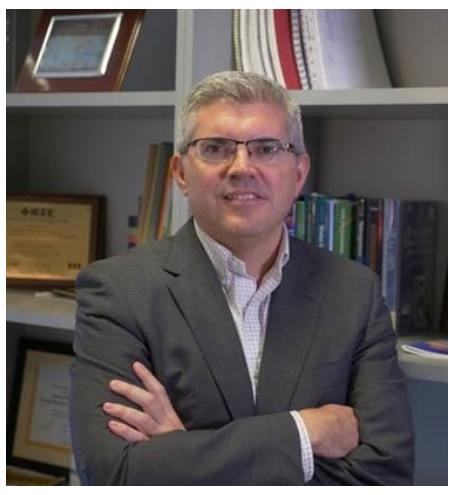

**Javier Gozalvez** received an electronics engineering degree from the Engineering School ENSEIRB (Bordeaux, France), and a PhD in mobile communications from the University of Strathclyde, Glasgow, U.K. Since October 2002, he is with UMH, where he is a Full Professor and Director of the UWICORE laboratory. At UWICORE, he leads research activities in the areas of vehicular networks, 5G and Beyond and industrial wireless networks. He is an elected member to the Board of Governors of the IEEE Vehicular Technology Society (VTS) since 2011, and served as its 2016-2017 President. He was an IEEE Distinguished Lecturer for the IEEE VTS, and currently serves as IEEE Distinguished Speaker. He is the Editor in Chief of the IEEE Vehicular Technology Magazine, and General Co-Chair for the IEEE Connected and Automated Vehicles Symposium 2020-2018, and IEEE VTC-Spring 2015.

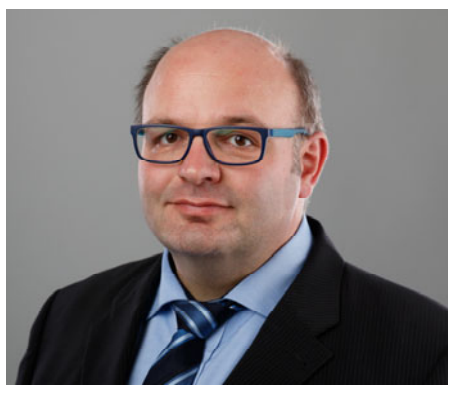

**Michael Reke** received a diploma degree in electronics engineering and a doctoral degree in computer science from RWTH Aachen University, Germany. From 2001 he was with VEMAC, Aachen, a company specialized in automotive electronics. From 2009 to 2017 he headed the research and development activities of VEMAC as CTO until in 2017 he was called to the FH Aachen, University of Applied Science, as a Professor for Vehicle Software. His research topics are software architectures for model-based development and automated driving.

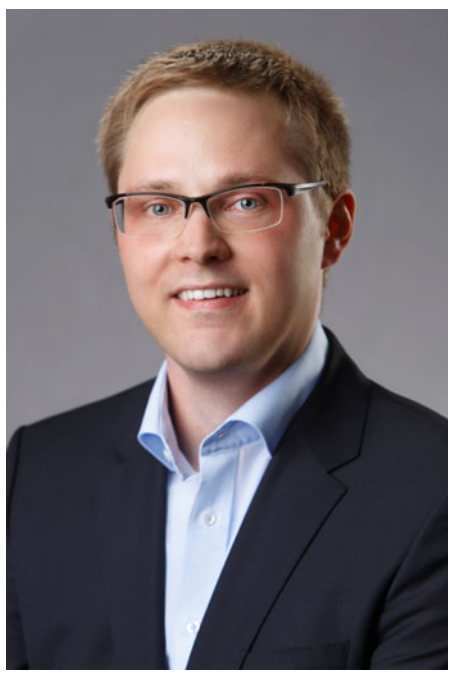

**Dominik Matheis** holds a MSc in Information Technology from the University of Applied Sciences in Mannheim (Germany). During his studies he was involved in multiple projects on identification and localization. He is currently a Senior Engineer at the Hyundai Motor Europe Technical Center (Germany), where he supports internal and EU-funded R&D projects on connected automated driving where he is responsible for map-based localization, AD vehicle setup including sensor HW/SW and localization for AD functions.

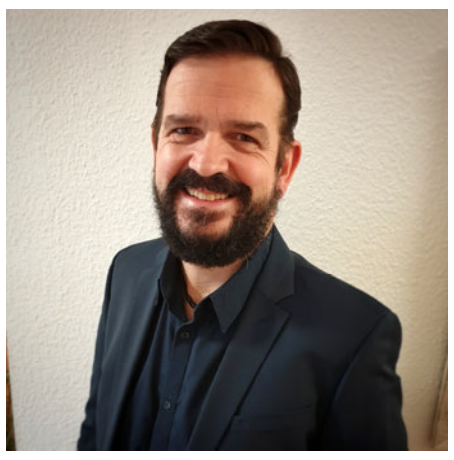

**Thomas Walter** holds an engineering degree in automation technology with focus on data systems technology from the University of applied science of Darmstadt (Germany). He worked more than 20 years at OEM level in the E/E & service development on data communication protocols (in-vehicle & off-board) and their standardization at ISO level. He currently leads the Advanced Safety Control Team of the Hyundai Motor Europe Technical Center (Germany) covering activities on connected automated driving and V2X.